\documentclass[11pt,sort&compress]{elsarticle}
\usepackage[paper=letterpaper,top=24mm, bottom=26mm, left=26mm, right=26mm]{geometry}
\makeatletter
\def\ps@pprintTitle{%
 \let\@oddhead\@empty
 \let\@evenhead\@empty
 \def\@oddfoot{\centerline{\thepage}}%
 \let\@evenfoot\@oddfoot}
\makeatother
\usepackage{amsmath}
\usepackage{accents}
\usepackage{amssymb}
\usepackage{mathrsfs}
\usepackage{mathtools}
\usepackage[LGR,T1]{fontenc}
\usepackage[latin1]{inputenc}
\let\oldbibliography\thebibliography
\renewcommand{\thebibliography}[1]{%
  \oldbibliography{#1}%
  \setlength{\itemsep}{1.4pt}%
}
\usepackage[cal=boondox,calscaled=1]{mathalfa}
\DeclareMathAlphabet{\bbvar}{U}{BOONDOX-ds}{m}{n}
\DeclareMathAlphabet{\bbgreek}{U}{bbold}{m}{n}
\usepackage[defaultsans,osfigures,scale=0.9]{opensans}
\usepackage{graphicx}
\usepackage{psfrag}
\usepackage{pstool}
\usepackage{caption}
\usepackage{graphicx}%
\newcommand{\hook}{\text{\large{$\lrcorner$}}}

\usepackage[colorlinks=true,
            linkcolor=blue,
            urlcolor=blue,
            citecolor=blue]{hyperref}
\usepackage[hyperref]{xcolor}
\definecolor{darkred}{rgb}{.95,.0,.0}

\newcommand{\di}{\mathrm{d}}
\usepackage{tensor}\newcommand{\ou}[3]{{#1}{}^{#2}{}_{#3}}
\newcommand{\uo}[3]{{#1}{}_{#2}{}^{#3}}

\newcommand{\I}{\mathrm{i}} %imaginaere Einheit
\newcommand{\E}{\mathrm{e}} %Euler Zahl
\newcommand{\CC}{\mathrm{cc.}} % komplex konjugiertes
\newcommand{\HC}{\mathrm{hc.}} % hermitsch konjugiertes
\newcommand{\C}{\mathbb{C}}
\newcommand{\N}{\mathbb{N}}
\newcommand{\R}{\mathbb{R}}

\newenvironment{subalign}{\subequations\align}{\endalign\endsubequations}
\newcommand{\eref}[1]{(\ref{#1})}

\newcommand{\bbwedge}{\reflectbox{\rotatebox[origin=c]{180}{\fontsize{10pt}{10pt}$\hspace{0.7pt}\bbvar{V}\hspace{0.7pt}$}}}

\newcommand{\AdS}{\mathrm{AdS}_{3}}
\renewcommand{\H}{\mathbb{H}_3}
\newcommand{\sh}{\operatorname{sh}}
\newcommand{\ch}{\operatorname{ch}}
\renewcommand{\th}{\operatorname{th}}
\newcommand\vpm{\mathbin{\vcenter{\hbox{
  \oalign{\hfil$\scriptstyle+$\hfil\cr
          \noalign{\kern-.3ex}
          $\scriptscriptstyle({-})$\cr}}}}}
\DeclareMathAlphabet{\sfit}{OT1}{fos}{sb}{it}
\DeclareMathAlphabet{\mathsf}{OT1}{fos}{sb}{n}

\definecolor{darkgreen}{rgb}{0.01, 0.75, 0.24}
\definecolor{darkblue}{rgb}{0.01, 0.24, 0.75}

\usepackage[multiple, flushmargin]{footmisc}

\usepackage{setspace}
\let\originalleft\left
\let\originalright\right
\renewcommand{\left}{\mathopen{}\mathclose\bgroup\originalleft}
\renewcommand{\right}{\aftergroup\egroup\originalright}
\newcommand{\Tr}{\operatorname{Tr}}
\begin{document}

\begin{abstract}
In this paper, we consider the bulk plus boundary phase space for three-dimensional gravity with negative cosmological constant for a particular choice of conformal boundary conditions: the conformal class of the induced metric at the boundary is kept fixed and the mean extrinsic curvature is constrained to be one. Such specific conformal boundary conditions define so-called Bryant surfaces, which can be classified completely in terms of holomorphic maps from Riemann surfaces into the spinor bundle. To study the observables and gauge symmetries of the resulting bulk plus boundary system, we introduce an extended phase space, where these holomorphic maps are now part of the gravitational bulk plus boundary phase space. The physical phase space is obtained by introducing two sets of Kac\,--\,Moody currents, which are constrained to vanish. The constraints are second-class and the corresponding Dirac bracket yields an infinite-dimensional deformation of the Heisenberg algebra for the spinor-valued surface charges. Finally, we compute the Poisson algebra among the generators of conformal diffeomorphisms and demonstrate that there is no central charge. Although the central charge vanishes and the boundary CFT is likely non-unitary, we will argue that a version of the Cardy formula still applies in this context, such that the entropy of the BTZ black hole can be derived from the degeneracy of the eigenstates of quasi-local energy.
\end{abstract}%

\title{Deformed Heisenberg charges in three-dimensional gravity}
%\title{Deformed gravHeisenberg algebra from Bryant surfaces in $\AdS$}
\author{Jeevan Chandra Namburi}
\author{\hspace{-0.5em}${}^{a, b}$ Wolfgang Wieland${}^{b}$}
\address{${}^a$Indian Institute of Science, Bangalore 560012, India\\${}^{a, b}$Perimeter Institute for Theoretical Physics\\31 Caroline Street North\\ Waterloo, ON N2L\,2Y5, Canada\\{\vspace{0.5em}\normalfont December 2019}
}
%\date{Winter 2017}
%\keywords{test}
\maketitle
\vspace{-1.2em}
\hypersetup{
  linkcolor=black,
  urlcolor=black,
  citecolor=black
}
{\tableofcontents}
%:
\hypersetup{
  linkcolor=darkred,
  urlcolor=darkred,
  citecolor=darkred
}
\begin{center}{\noindent\rule{\linewidth}{0.4pt}}\end{center}\newpage
\section{Introduction}
\noindent  How do we characterise a solution to Einstein's equations on the gravitational phase space? Consider, for example, the Kerr metric, which is determined by the mass $M$ and the spin $J\leq M^2$ of the black hole. If appropriate gauge-fixing, parity, and falloff conditions for the metric at infinity are satisfied, there are ten conserved Poincaré charges at spacelike infinity \cite{adm,Beig:1987zz,Regge:1974zd}, and  every pair $(J,M)$, $J\neq 0$ represents an entire 10-dimensional submanifold on the ADM phase space for asymptotically flat boundary conditions. On the radiative phase space \cite{Bondi21,Sachs103,AshtekarNullInfinity,Strominger:2017zoo}, which describes the radiative modes at future (past) null infinity, the situation is a little more difficult, because (i) future (past) null infinity is not a complete Cauchy hypersurface, and (ii) there are infinitely many ways to embed the Poincaré group into the asymptotic symmetry group of BMS transformations, which is infinite-dimensional. At the quantum level, these infrared ambiguities \cite{AshtekarNullInfinity,Ashtekar:2018lor,Ashtekar:1981sf,Strominger:2017zoo} give rise to infinitely many unitarily inequivalent Fock representations. Choosing a radiative Fock vacuum amounts to choosing a specific Poincaré subgroup at future (past) null infinity \cite{AshtekarNullInfinity,Ashtekar:2018lor,Ashtekar:1981sf}, and there are infinitely many unitarily inequivalent such choices. If one insists that superpositions of different such unitarily inequivalent radiative vacua should be realised in nature, and also takes into account that BMS translations define canonical transformation on the radiative phase space \cite{AshtekarNullInfinity}, one is naturally led to the idea that the BMS transformations should arise from a non-unitary boundary field theory. {Such theories still admit a positive-definite inner product, and there are no negative probabilities, but the condition is dropped that the Hamiltonian (e.g.\ a generic BMS supertranslation) preserves the different unitarily inequivalent Fock spaces. %We may expect, therefore, that the boundary theory should live on a direct sum of orthogonal Fock spaces, $\mathcal{H}=\int_\oplus d\alpha\,\mathcal{F}_\alpha$.} 

In the following, we will collect evidence in favour of this scenario by considering three-dimensional Euclidean gravity (with negative cosmological constant). In addition, we will work on a quasi-local phase space, where the gravitational field is put in a box with boundaries at finite distance \cite{Szabados:2004vb,Wald:1999wa,Andrade:2015fna,Harlow:2019yfa,Chandrasekaran:2019ewn,Wieland:2019hkz,Freidel:2019ees}. The topology of the boundary is fixed: the boundary is an infinite cylinder $\R\times S^1$, which serves as a toy model for future (past) null infinity \cite{Barnich:2011mi,Barnich:2006av,Barnich:2014kra,Barnich:2015uva,Afshar:2016kjj}. At this cylindrical boundary, we then choose specific conformal boundary conditions \cite{Witten:2018lgb}. Although there is just one single solution of Einstein's equations in the interior, namely empty $\AdS$, the physical phase space of the bulk plus boundary system turns out to be \emph{infinite-dimensional}. The physical origin of this vacuum degeneracy has to do with the introduction of the boundary, which turns otherwise unphysical gauge directions into actual physical boundary degrees of freedom \cite{Carlip:1996yb,Rovelli:2013fga,Gomes:2018shn}.\footnote{
There are infinitely many ways to embed the abstract boundary $\R\times S^1$ into $\AdS$, and each of these different embeddings represents a physically distinguished configuration of the bulk plus boundary field theory.} Working in the first-order connection representation, we will then see that these inequivalent boundary configurations can be characterised by a boundary boundary spinor that is coupled to the spin connection in the bulk. The underlying boundary field theory has, however, a number of unexpected features: the Virasoro algebra, which is the Euclidean analogue of the algebra of BMS transformations, has no anomaly (the central charge vanishes at the level of classical Poisson brackets), and the kinetic term for the boundary fields is not positive-definite. In fact, the boundary action defines a version of the $\beta$-$\gamma$ ghosts of superstring theory with a quartic self interaction  $V(\beta,\gamma)\propto (|\beta|^2+|\gamma|^2)^2$, which suggests that the resulting boundary field theory defines a non-unitary CFT. A central charge may reappear upon quantisation, but it would have the wrong $\hbar$ dependence that would be required to derive the entropy of a BTZ black hole from the Cardy formula, see \cite{Carlip:2001kk}, where a similar issue arises for the Liouville boundary CFT.\vspace{-0.6em} 

\paragraph{- Outline} The paper is organised as follows. First of all (section 2), we introduce a specific class of conformal boundary conditions, such that the abstract boundary $\R\times S^1$ is mapped into a constant mean curvature one hypersurface (CMC-1 in units of $\ell=\sqrt{-1/\Lambda}$), which is embedded into three-dimensional hyperbolic space $\H$. We then show how these specific conformal boundary conditions translate into a holomorphicity condition for an $SU(2)$ spinor $\xi^A\equiv |\xi\rangle\in\C^2$, which is intrinsic to the boundary. The squared $SU(2)$ norm $\|\xi\|^2=\langle \xi|\xi\rangle$ of this boundary spinor determines the conformal factor, which relates the pull-back of the physical metric in the interior to the auxiliary metric at the boundary. The relation between the boundary spinor $\xi^A$ and the embedding variables is provided by the three-vector $\vec{n} = \langle\xi|\vec{\sigma}|\xi\rangle$, which defines the internal normal to the boundary: $\vec{n}\equiv n^i=\ou{e}{i}{a}n^a$, where $\ou{e}{i}{a}$ denotes the co-triad in the interior. To clarify the geometry of the problem, we consider then a particular class of such CMC-1 boundaries, namely \emph{Bryant's curved catenoid cousins} \cite{AST_1987__154-155__321_0,Bobenko2009}, and parametrise the solution space of the boundary field theory in terms of holomorphic maps (see section 3). Next, we add the appropriate counter  terms to the triadic Palatini action such that the Einstein equations in the interior and the additional conformal boundary conditions both follow from the saddle point equations of the coupled bulk plus boundary action. Section 4 deals with the quasi-local Hamiltonian analysis. We will introduce an extended gravitational phase space, and identify the gauge transformations (small diffeomorphisms and $SU(2)$ frame rotations) of the extend bulk plus boundary system \cite{Donnelly:2016auv,Freidel:2018pbr,Freidel:2019ees,Meneses:2019bok,Oliveri:2019gvm,DePaoli:2018erh}. The commutation relations between the Laurent modes of the boundary spinor $\xi^A$ are defined via the Dirac bracket, which yields a deformation of the infinite-dimensional Heisenberg algebra. The strength of the deformation is determined by the cosmological constant. If the cosmological constant vanishes, the deformation disappears. Finally, we turn to quantum gravity and explain under which assumptions the boundary conformal field theory could provide a concrete realisation of Strominger's proposal \cite{Strominger:1997eq} for a microscopic derivation of black hole entropy from the degeneracy of the eigenstates of quasi-local energy.\vspace{-0.6em}
 
\paragraph{- Notation} Our conventions are the following: $a,b,c,\dots$ are abstract tensor indices, and we will use them without any distinction for both tensor fields in space time and for tensor fields that are intrinsic to the two-dimensional boundary $\mathcal{B}$ of the three-dimensional cylinder $\mathcal{M}\simeq \R\times\Sigma$. Two-dimensional spinor indices $A,B,C,\dots$ carry a representation of $SL(2,\C)$, the complex conjugate representation is denoted by primed indices $A',B',C',\dots$. The skew-symmetric and $SL(2,\C)$ invariant $\epsilon$-tensor provides a map between covariant and contravariant such spinors, i.e.\ $\xi^A=\epsilon^{AB}\xi_B,\xi_B=\xi^A\epsilon_{AB}$.  Round (square) brackets surrounding indices $A_1,A_2,\dots$ denote total (anti)symmetrisation, i.e.\ $2\omega_{(AB)}=\omega_{AB}+\omega_{BA}$. In addition, there is also an $SU(2)$ invariant inner product, $\langle\eta|\xi\rangle=\delta_{AA'}\bar{\eta}^{A'}\xi^A$, which allows us to define the Hermitian conjugate $\xi^\dagger_A =\delta_{AA'}\bar{\xi}^{A'}$, $\xi_\dagger^A=\epsilon^{AB}\xi^\dagger_B$. An element $U\in SU(2)$ can be then identified with those tensors $\ou{U}{A}{B}$ in the spin $(\tfrac{1}{2})\otimes({\tfrac{1}{2}})^\ast$ representation that preserve $\delta_{AA'}$, i.e.\ $\delta_{AA'}=\delta_{BB'}\ou{U}{B}{A}\ou{\bar{U}}{B'}{A'}$. This notation is convenient for us, because it allows us to make sense of sums and differences of group elements (we are implicitly working on the universal enveloping algebra, as in equation \eref{Cdef} below). Finally, let us also mention that $\ou{\sigma}{A}{Bi}\equiv\sigma_i$ are the usual Pauli matrices and $\tau_i=1/(2\I)\sigma_i$ is the corresponding basis in $\mathfrak{su}(2)$ that satisfies $[\tau_i,\tau_j]=\uo{\epsilon}{ij}{k}\tau_k$.
\section{Bulk plus boundary field theory for conformal boundary conditions}\label{sec2}
\subsection{Conformal boundary conditions in $AdS_{3}$}
\noindent In the quasi-local covariant phase space approach boundary conditions on a $t=\mathrm{const}.$ initial hypersurface $\Sigma$ have a slightly different ontological status than those for the timelike\footnote{The distinction between timelike and spacelike hypersurfaces is meaningless in Euclidean gravity, but we can always work with an Euclidean $t$-coordinate, with respect to which the equations of motion of the bulk plus boundary field theory can be cast into a standard Hamiltonian form.} portion $\mathcal{B}\subset\partial\mathcal{M}$ of the boundary: different boundary conditions on $\mathcal{B}$ select different Hamiltonians on a extended phase space $\mathcal{P}_\Sigma$ of the bulk plus boundary system \cite{Wieland:2017zkf,Donnelly:2016auv,Harlow:2019yfa}, and the  boundary conditions on $\mathcal{B}$ translate into external sources (background fields or $c$-numbers) that parametrise the possible (time dependent) Hamiltonians on $\mathcal{P}_\Sigma$. %Different boundary conditions on $\mathcal{B}$ amount to altogether different Hamiltonians on the extended phase space $\mathcal{P}_\Sigma$ representing altogether different bulk plus boundary field theories. 
In three dimensions, this procedure is comparably easy to understand, because once we fix the gauge conditions along $\mathcal{B}$ there is no additional free data left, since three-dimensional gravity is topological. %Nevertheless, three-dimensional gravity still provides valuable insights into the classical and quantum aspects of the problem: although there are no propagating degrees of freedom in the interior of spacetime there still is an infinite tower of boundary degrees of freedom that describe how the boundary of an initial hypersurface is embedded into the physical spacetime geometry. The same boundary degrees of freedom also appear in higher dimensions, where there are indications that they are the relevant microscopic degrees of freedom that are responsible for black hole entropy.
%In the following, we restrict ourselves to three-dimensional Euclidean gravity with a negative cosmological constant
%\begin{equation}
%\Lambda = -\frac{1}{\ell^2}.
%\end{equation}
%In addition, we are following a quasi-local \cite{Szabados:2004vb} phase space approach, where we treat the gravitational bulk plus boundary degrees of freedom in a finite box as a Hamiltonian system on a phase space $\mathcal{P}_\Sigma$. 

The boundary conditions on $\mathcal{B}$ determine how the boundary $\partial\Sigma\simeq S^1$ of the initial hypersurface $\Sigma$, which has the topology of the unit disk $\big\{z\in \C\big||z|\leq 1\big\}$, extends into a world tube $\mathcal{B}\simeq \R\times S^1$, which is embedded into spacetime. A particular simple possibility to determine such an embedding is given by the following conformal boundary conditions: the basic idea is to only fix the boundary metric up to conformal transformations,
\begin{equation}
\varphi^\ast_{\mathcal{B}} g_{ab} =: h_{ab} \in [q_{ab}]\Leftrightarrow \Omega:\mathcal{B}\rightarrow\R : h_{ab} = \Omega^{-2} q_{ab}.
\end{equation}
Similar boundary conditions can be used  in 3+1 dimensions, where the conformal two-structure of the light cone  determines the two-radiative modes at the full non-perturbative level \cite{Sachs:1962zzb,InvernoStachelConformal}. The conformal factor $\Omega$, on the other hand, is unconstrained. Instead, we freeze its conjugate momentum, which is the trace of the extrinsic curvature\footnote{On the ADM phase space, we can always choose a polarisation such that $\log\Omega$ is canonically conjugate to the trace of the extrinsic curvature.}
\begin{equation}
K= h^{ab}K_{ab},
\end{equation}
where $K_{ab}=\uo{h}{a}{c}\nabla_c n_b$ is the extrinsic curvature tensor and $n^a:g_{ab}n^an^b=1$ is the outwardly oriented normal to the boundary. We choose, therefore, the following conformal boundary conditions,
\begin{equation}
\delta[K] =0,\qquad \delta h_{ab} \propto q_{ab},\qquad \delta [h_{ab}] = 0.\label{bndrycond}
\end{equation}
Since $\delta[K]$ vanishes, we have to choose a specific value for $K$. We will see in the following that
\begin{equation}
K = \frac{2}{\ell}\label{bndrycond2}
\end{equation}
is preferred geometrically, because it selects specific \emph{Bryant surfaces} that are in one-to-one correspondence to holomorphic maps from the punctured complex plane $\C-\{0\}$ into the spinor bundle over hyperbolic space \cite{AST_1987__154-155__321_0,Bobenko2009}.

\subsection{Bulk plus boundary field equations}
\noindent The action in the interior of the cylinder $\mathcal{M}\simeq\R\times\Sigma$ is given by the usual triadic Palatini action,
\begin{equation}
S_{\mathcal{M}}[e,\omega] = - \frac{1}{8\pi G}\int_{\mathcal{M}}\Big[e_i\wedge F^i[\omega]+\frac{\Lambda}{6}\epsilon_{ilm}e^i\wedge e^l\wedge e^m\Big],\label{bulkactn}
\end{equation}
where $\omega^i$ denotes an $SU(2)$ connection with curvature $F^i[\omega]=\di\omega^i+\frac{1}{2}\ou{\epsilon}{i}{jk}\omega^k\wedge \omega^k$ and  $e^i$ is the co-triad. The metric tensor is the composite field
\begin{equation}
g_{ab}=\delta_{ij}\ou{e}{i}{a}\ou{e}{j}{b}.
\end{equation}
If the torsionless condition $T^i = \di e^i +\ou{\epsilon}{i}{lm}\omega^l\wedge e^m = 0$ is satisfied, the action \eref{bulkactn} reduces to the usual Einstein\,--\,Hilbert action
\begin{equation}
S_{\mathrm{EH}}[g] = \frac{1}{16\pi G}\int_{\mathcal{M}}d^3v_g\big(R[g]-2\Lambda\big),
\end{equation}
where $d^3 v_g = 1/3!\,\epsilon_{ilk}e^i\wedge e^l\wedge e^k$ is the metrical volume element and $R[g]$ denotes the Ricci scalar.

The torsionless condition is satisfied at the stationary points of the action. A generic such variation yields a boundary term $1/(8\pi G)\oint_{\mathcal{B}} e_i\wedge\delta\omega^i$. To make the action for the conformal boundary conditions \eref{bndrycond} functionally differentiable, we have to cancel this boundary variation by the addition of an appropriate counter term. Since the reminder $1/(8\pi G)\oint_{\mathcal{B}} e_i\wedge\delta\omega^i$ of the $\omega^i$-variation is linear in the connection, we will construct such a boundary term from the covariant derivative, which acts \emph{linearly} on an $SU(2)$ boundary spinor $\iota^A$,
\begin{equation}
\nabla_a\iota^A = \partial_a
\iota^A + \ou{\tau}{A}{Bi}\ou{\omega}{i}{a}\iota^B,
\end{equation}
where $\partial_a$ is a flat reference connection and $\tau_i = 1/(2\I) \sigma_i$ are the $\mathfrak{su}(2)$ generators.  The task ahead is  to find such a boundary term for the conformal boundary conditions (\ref{bndrycond}, \ref{bndrycond2}) and add it to the action. %The resulting boundary action will define a conformal boundary field theory for an $SU(2)$ boundary spinor coupled to the spin connection in the bulk. %At the discrete level, such boundary spinors have appeared recently from the semi-classical and saddle point analyses of the loop gravity and spinfoam approaches to quantum gravity: in loop quantum gravity, the quantum states of geometry consist of superpositions of gravitational Wilson loops. If we include boundaries and restrict our attention to the gravitational degrees of freedom that are confined within such a boundary, some of the gravitational Wilson loops will intersect the boundary at a puncture, where they excite a spinor-valued surface charge \cite{Barrett:2009gg,Wieland:2017cmf,wieland:nulldefects,Wieland:2017zkf,Ashtekar:2000eq,Dittrich:2018xuk,Dittrich:2017rvb}. The dynamics of these surface charges along the boundary is induced by the evaluation of the boundary states against the spinfoam amplitudes. In three dimensions and for vanishing cosmological constant, this procedure defines a large class of dual statistical models \cite{Bonzom:2015ans,Dittrich:2018xuk,Dittrich:2017hnl,Dittrich:2017rvb}.
To impose the boundary conditions \eref{bndrycond}  in terms of such surface spinors, let us first write the extrinsic curvature in terms of a spin frame at the boundary. Such a spin frame can be defined by a single and normalised $SU(2)$ spinor $\iota^A:\delta_{AA'}\iota^A\bar{\iota}^{A'}=1$, which immediately defines a second and linearly independent and orthogonal spinor $o^A$,
\begin{equation}
o^A :=\iota_\dagger^A \equiv \epsilon^{AB}\iota^\dagger_B = \epsilon^{AB}\delta_{BB'}\bar{\iota}^{B'},
\end{equation}
where $\delta_{AA'}$ denotes the $SU(2)$ invariant Hermitian metric and $\epsilon^{AB}$ is the skew symmetric $\epsilon$-tensor.\footnote{The $SU(2)$ and $SL(2,\C)$ spinor indices are raised and lowered with respect to the skew symmetric and $SL(2,\C)$-invariant $\epsilon$-tensors, e.g.\ $\xi_A = \epsilon_{BA}\xi^B$, $\xi^A = \epsilon^{AB}\xi_B$, and $\epsilon^{AB}\epsilon_{AB}=2$, see \cite{penroserindler}.} Given the spin dyad $(o^A,\iota^A)$, we can then immediately construct a corresponding internal triad,%\footnote{The map between the two columns defines an isometry $X^i\longrightarrow X^{AB}=X^{BA}$ for inner products $X_{AB}Y^{AB}=X_iY^i$.}
\begin{subalign}
v^i  &= \frac{\I}{\sqrt{2}}\uo{\sigma}{AB}{i}v^{AB}, & \hspace{-3em}v^{AB}&=-\I\sqrt{2}\, o^{(A}\iota^{B)},\label{iso1}\\
w^i  &= \frac{\I}{\sqrt{2}}\uo{\sigma}{AB}{i}w^{AB},&\hspace{-3em} w^{AB}&=+\I o^Ao^B,\label{iso2}\\
\bar{w}^i  &= \frac{\I}{\sqrt{2}}\uo{\sigma}{AB}{i}w^{AB},&\hspace{-3em}\bar{w}^{AB}&=-\I\iota^A\iota^B\label{iso3}.
\end{subalign}
It is easy to check that $\bar{w}_iw^i=v_iv^i =1$, while all other contractions vanish (internal indices are raised and lowered with the flat internal metric $\delta_{ij}$). In addition, $\bar{w}^i$ is the complex conjugate of $w^i$ and $\ou{\sigma}{A}{Bi}=\epsilon^{AC}\sigma_{CBi}$ are the usual Pauli matrices. Consider then a smooth section $(o^A,\iota^A)$ of the associated frame bundle in some neighbourhood of $\partial\mathcal{M}$. Given the triad $\uo{e}{i}{a}$, we can now introduce an associate spacetime triad $(v^a,w^a,\bar{w}^a)$, where e.g.\ $v^a = \uo{e}{i}{a}v^i$. The $SU(2)$ covariant derivative annihilates the Pauli matrices. If the torsionless condition is satisfied, it also annihilates  the triad\footnote{On-shell, the covariant derivative satisfies $\nabla_a\uo{e}{i}{b} = \partial_a\uo{e}{i}{b}+\uo{\epsilon}{il}{m}\ou{\omega}{l}{a}\uo{e}{m}{b}+\ou{\Gamma}{b}{ca}\uo{e}{i}{c}=0$, where  $\ou{\Gamma}{a}{bc}$ are the Christoffel symbols for the metric $g_{ab}=\delta_{ij}\ou{e}{i}{a}\ou{e}{j}{b}$.} $\ou{e}{i}{a}$, and it is then easy to see that the expansion $\vartheta$ and the twist $\omega$ of $v^a$ reduce to the following complex-valued spin coefficient,
\begin{equation}
\frac{1}{2}\big(\vartheta-\I\omega\big):=w^a\bar{w}^b\nabla_av_b=-\sqrt{2}\,\iota_A\iota_Bw^a\nabla_a\big(o^{A}\iota^{B}\big)=\sqrt{2}\,\iota_Aw^a\nabla_a\iota^A,\label{expdef}
\end{equation}
where we were using the normalisation of the dyadic spinor basis: $\iota_A\iota^A = \epsilon_{AB}\iota^A\iota^B=0$ and $o_A\iota^A=-\iota_Ao^A=1$.

If we extend $n^a$ into a surface forming vector field in the neighbourhood of $\mathcal{B}$, and align $n^a$ with the vector field $v^a=\uo{e}{i}{a}v^i$, we immediately see that the twist $\omega$ of $v^a$ must vanish (since $n^a$ is surface forming), while the conformal boundary condition $K=2/\ell$ translates into a condition for the $\iota_A\nabla\iota^A$ spin coefficient, 
\begin{equation}
\iota_A w^a\nabla_a\iota^A = \frac{1}{\sqrt{2}}\frac{1}{\ell}.\label{bndrycond3}
\end{equation}
The conformal boundary conditions impose constraints on \emph{both} the extrinsic curvature and the induced metric. Having expressed the boundary condition $K=2/\ell$ in terms of the spin coefficients, we have done only one half of the job. We must now turn to the boundary condition for the off-diagonal components of the induced metric and rewrite it as a boundary condition for the spin frame at the boundary. The induced metric is
\begin{equation}
\varphi^\ast_{\mathcal{B}}	 g_{ab} = \Omega^{-2} q_{ab} = 2\Omega^{-2} m_{(a}\bar{m}_{b)},
\end{equation}
where $m_a\in \Omega^1(\mathcal{B}:\C)$ defines a reference dyad on the boundary and $\Omega$ is the conformal factor.\footnote{The dyadic one-form $m_a$ is a $c$-number on phase space, $\delta[m_a]=0$.} This condition can be easily translated into the spinor calculus. The basic idea is to align the one-form $m_a$ with a spinor $\xi^A\xi^B$ such that the $SU(2)$ norm of $\xi^A$ determines the conformal factor,\footnote{Notice that both $m_a$ and $\xi_A\xi_A$ are null: $m_am^a=0=\xi_A\xi_B\xi^A\xi^B$.}
\begin{equation}
\varphi^\ast_{\mathcal{B}} e^i =4\pi G\,\Big(\frac{1}{\sqrt{2}}\xi^A\xi^B\uo{\sigma}{AB}{i}m+\CC\Big),\label{glucond}
\end{equation}
see \cite{Wieland:2018ymr}. In fact, it is easy to check that the conformal factor turns into a composite field,
\begin{equation}
\Omega^{-1} = 4\pi G \|\xi\|^2\equiv 4\pi G\,\delta_{AA'}\xi^A\bar{\xi}^{A'},\label{Omxi}
\end{equation}
where $4\pi G = \ell_P$ denotes the Planck length in three spacetime dimension (in units of $\hbar = 1$). The Planck length has been introduced for dimensional reasons only, in quantum gravity, on the other hand, equation \eref{Omxi} defines the most natural normalisation: for vanishing cosmological constant, $\|\xi\|^2$ turns into an ordinary number operator, whose spectrum is $\N+\frac{1}{2}$, see \cite{Wieland:2018ymr}. %The boundary spinor $\xi^A$ will be the fundamental configuration variable for the boundary field theory. 

At the boundary, there is a natural torsionless $SU(2)\times U(1)$ covariant derivative $D_a$. It annihilates the dyadic one-forms $m_a$, which are uncharged under $SU(2)$, and it satisfies the two-dimensional torsionless condition,
\begin{equation}
D m = \tensor[^2]{\di}{}m + \I\,\Gamma\wedge m=0,\label{Btors}
\end{equation}
where $\Gamma\in\Omega^1(\mathcal{B})$ is the $U(1)$ boundary spin connection and $\tensor[^2]{\di}{}$ is the exterior derivative on $\mathcal{B}$. By adding the spin connection from the bulk, the $U(1)$ covariant derivative, which is defined by $\Gamma_a$, extends naturally to an $SU(2)\times U(1)$ covariant derivative, which acts on the boundary spinors via
\begin{equation}
D\xi^A = \tensor[^2]{\di}{}\xi^A -\frac{\I}{2}\Gamma_a\xi^A + \ou{\tau}{A}{Bi}(\varphi^\ast_{\mathcal{B}}\omega^i)\xi^B.\label{Dxi}
\end{equation}
The torsionless condition in the bulk imposes now a constraint on this derivative. Since the exterior derivative commutes with the pull-back, we easily find
\begin{equation}
0=\varphi^\ast_{\mathcal{B}}\nabla e^i = D\big(\varphi^\ast_{\mathcal{B}}e^i\big) = 4\pi G\Big(\sqrt{2}\,\xi^{(A}\big(D\xi^{B)}\big)\uo{\sigma}{AB}{i}\wedge m+\CC\Big).\label{2dtors}
\end{equation}
Since $\xi^A\xi^B$, $\xi_\dagger^A\xi_\dagger^B$ and $\xi^{(A}\xi^{B}_\dagger{}^{)}$ are linearly independent and define a complexified basis in the $SU(2)$ Lie algebra, we conclude  that the pull-back of the torsionless condition vanishes if and only if
\begin{equation}
m\wedge D\xi^A\propto\xi_\dagger^A=\ou{\delta}{A}{A'}\bar{\xi}^{A'}.\label{Dxi2}
\end{equation}
The proportionality between the right hand side and the left hand side is determined by the extrinsic curvature. To establish the relation between the extrinsic curvature and $D\xi^A$, we  introduce a normalised spin frame $(\iota^A,o^A)$, which is aligned to $\xi^A$, such that we can infer the extrinsic curvature from the spin coefficient $\iota_A\nabla\iota^A$, see   \eref{expdef}. Consider, therefore, the following spin frame at the boundary,
\begin{equation}
\iota^A = \frac{\xi^A}{\|\xi\|},\quad o^A =\frac{\ou{\delta}{A}{A'}\bar{\xi}^{A'}}{\|\xi\|}.
\end{equation}
There are now two associate bases in $T^\ast\mathcal{B}_\C$, namely $(m^a,\bar{m}^a)$, which is defined as the basis dual to the dyadic one-forms $(m_a,\bar{m}_a)$,\footnote{i.e.\ $m^am_a=0, \bar{m}^am_a=0$.} and $(w^a,\bar{w}^a)$, which is induced from the bulk, see (\ref{iso1}, \ref{iso2}, \ref{iso3}). The two bases are related by the conformal factor,
\begin{equation}
w^a = \Omega m^a,\label{wmdef}
\end{equation}
which is determined from the $SU(2)$ norm \eref{Omxi} of the boundary spinor. Going back to the definition of the extrinsic curvature in terms of the spin coefficients, i.e.\ equation \eref{expdef}, we can now finally determine the relation \eref{Dxi2} between $m\wedge D\xi^A$ and $\xi^A_\dagger$,
\begin{equation}
m\wedge D\xi^A = - \frac{1}{\|\xi\|^2}\xi^A_\dagger \xi_B m\wedge D\xi^B = -\xi^A_\dagger m\wedge \iota_BD\iota^B=-\frac{1}{2\sqrt{2}}\,\Omega^{-1}\big(\vartheta-\I\omega\big)\xi_\dagger^A m\wedge\bar{m}.
\end{equation}
The conformal boundary condition (\ref{bndrycond2}) turns, therefore, into the following holomorphicity condition for the boundary  spinor $\xi^A$,
\begin{equation}
K=\frac{2}{\ell}\Leftrightarrow\,{m}^aD_a\xi^A = - \frac{2\sqrt{2}\,\pi G}{\ell}\|\xi\|^2\ou{\delta}{A}{A'}\xi^{A'}.\label{EOM1}
\end{equation}
In the following, we will treat this boundary condition as a dynamical field equation, which is derived from the coupled bulk plus boundary action.  
\subsection{Bulk plus boundary action}
\noindent Now that we have identified the boundary field equations \eref{EOM1} that impose the conformal boundary condition $K=2/\ell$, it is immediate to infer the corresponding bulk plus boundary action. In fact, the action for the coupled bulk plus boundary system is given by the usual triadic Palatini action in the interior and the action for a two-dimensional field theory at the boundary,
\begin{align}\nonumber
S[e,\omega|\xi] = - \frac{1}{8\pi G}\int_{\mathcal{M}}&\bigg[e_i\wedge F^i[\omega]+\frac{\Lambda}{6}\epsilon_{ilm}e^i\wedge e^l\wedge e^m\bigg]+\\
&+\frac{\I}{\sqrt{2}}\int_{\mathcal{B}}\bigg[\xi_A m\wedge D\xi^A-\bar{\xi}_{A'} \bar{m}\wedge D\bar{\xi}^{A'}-\frac{2\sqrt{2}\,\pi G}{\ell}m\wedge\bar{m}\|\xi\|^4\bigg],\label{actndef}
\end{align}
where the quartic potential $\|\xi\|^4 = (\delta_{AA'}\xi^A\bar{\xi}^{A'})^2 $ is built from the $SU(2)$ invariant Hermitian norm and $D$  denotes the $SU(2)\times U(1)$ boundary covariant derivative \eref{Dxi}. The equations of motion (EOM) in the interior are the three-dimensional Einstein equations plus the torsionless condition,
\begin{subalign}
T^i &= \nabla e^i = \di e^i +\ou{\epsilon}{i}{lm}\omega^l\wedge e^m =0,\label{TEOM}\\
F^i &= \di\omega^i + \tfrac{1}{2}\ou{\epsilon}{i}{lm}\omega^l\wedge \omega^m = -\frac{\Lambda}{2}\ou{\epsilon}{i}{lm}e^l\wedge e^m.\label{FEOM}
\end{subalign}
The boundary conditions $K=2/\ell$ along the cylindrical boundary follow as an additional boundary equation of motion from the variation of the action with respect to $\xi^A$ (resp.\ $\bar{\xi}^{A'}$). In fact,
\begin{align}\nonumber
\delta_\xi\big[S[e,\omega|\xi]\big] = \I\sqrt{2}\int_{\mathcal{B}}&\Big[\delta\xi_A\Big(m\wedge D\xi^A+\frac{2\sqrt{2}\pi G}{\ell}m\wedge\bar{m}\|\xi\|^2\ou{\delta}{A}{A'}\bar{\xi}^{A'}\Big)-\CC\Big]+\\
&\hspace{5em}+\frac{\I}{\sqrt{2}}\int_{\partial\mathcal{B}}\big(m\,\xi_A\delta\xi^A-\CC\big).\label{dxivar}
\end{align}
The first term imposes the boundary condition $K=2/\ell$, while the one-dimensional integrals at the one-dimensional corners $\partial\mathcal{B}=\partial\Sigma_{+}^{-1}\cup\partial\Sigma_-^{-1}$ will add a corner term to the pre-symplectic potential of the bulk plus boundary field theory, see \eref{thetadef1}.

Finally, there is also the \emph{gluing condition} \eref{glucond}, which is satisfied at the stationary points of the coupled bulk plus boundary action. This additional gluing condition follows from the $\omega^i$-variation of the coupled bulk plus boundary action. A short calculation gives,
\begin{align}\nonumber
\delta_\omega\big[S[e,\omega|\xi]\big]  = \frac{1}{8\pi G}&\int_{\mathcal{M}}T_i\wedge\delta\omega^i +\frac{1}{8\pi G}\int_{\mathcal{B}}\Big[e_i-4\pi G\Big(\frac{1}{\sqrt{2}}\xi_A\xi_B\ou{\sigma}{AB}{i}m+\CC\Big)\Big]\wedge\delta\omega^i+\\
&+ \int_{\Sigma_-}^{\Sigma^+} e_i\wedge\delta\omega^i.
\end{align}
The first line vanishes as an equation of motion: $T^i=0$ is the torsionless condition \eref{TEOM}, and the second term vanishes provided the gluing conditions \eref{glucond} are satisfied. The two boundary integrals in the second line define the contribution to the pre-symplectic potential from the interior, see \eref{thetadef1}.

The boundary equations of motion \eref{EOM1} can be simplified by introducing the $SL(2,\C)\times U(1)$ covariant derivative with respect to the Euclidean $\AdS$ connection,
\begin{equation}
\mathcal{D}\xi^A = D\xi^A +\frac{1}{2\ell}\ou{\sigma}{A}{Bi}(\varphi^\ast_{\mathcal{B}}e^i)\xi^B,\label{AdSD}
\end{equation}
The \emph{gluing condition} \eref{glucond} implies ${\sigma}_{ABi}\varphi^\ast_{\mathcal{B}}e^i = -4\pi G\sqrt{2} (\xi_A\xi_B m - \xi^\dagger_A\xi^\dagger_B\bar{m})$ such that $\mathcal{D}\xi^A = D\xi^A +2\sqrt{2}\pi G/\ell \|\xi\|^2\xi^A_\dagger\bar{m}$. The boundary equation of motion \eref{EOM1} reduces, therefore, to the simple holomorphicity condition
\begin{equation}
m^a\mathcal{D}_a\xi^A = 0\label{EOM2}
\end{equation}
for the boundary spinor $\xi^A$, where $\mathcal{D}_a$ is the $SL(2,\C)\times U(1)$ covariant derivative \eref{AdSD}.

\section{Solution space, curved catenoids, deformed Gauss law}\label{sec3}
\subsection{Particular solution: the $AdS_{3}$ catenoid}
\noindent To clarify the geometry of the system, let us consider first a particular solution of the bulk plus boundary field equations. The goal is, in other words, to find a diffeomorphism $\varphi$ that maps the solid cylinder\footnote{The two-dimensional disk $\Sigma$ is bounded by a circle $\partial\Sigma \simeq S^1$.} $\R\times\Sigma$ into Euclidean $\AdS$ (i.e.\ three-dimensional hyperbolic space $\H$ with cosmological constant $\Lambda =-1/\ell^2$) such that the trace of the extrinsic curvature of the boundary $\mathcal{B}=\varphi(\R\times\partial S^1)$ satisfies the constraint
\begin{equation}
K=\frac{2}{\ell}.\label{Kdef}
\end{equation}
To find an explicit example of such a \emph{Bryant surface} \cite{AST_1987__154-155__321_0}, we will work with cylindrical $\H$ coordinates $(\rho,\varphi,\eta)$. In these coordinates, the $\H$ line element is given by
 \begin{equation}
ds^2 = \ell^2 \big(\di\rho^2 + \sh^2\rho\,\di\varphi^2 +\ch^2\eta\,\di\eta^2\big).\label{gdef}
\end{equation}

The trace of the extrinsic curvature $K=h^{ab}K_{ab}$ is the three-divergence of the normal vector to the boundary, i.e.\ $K=h^{ab}\nabla_an_b=\nabla_an^a$, where $\nabla_a$ denotes the torsionless and metric compatible derivative in the bulk. To satisfy \eref{Kdef} consider then the following ansatz for the normalised vector field $n^a$,
\begin{equation}
n^a = N(\rho)\,\varepsilon^{abc}\,\partial_b\varphi\,\partial_c\big(\eta-f(\rho)\big),\label{ndef}
\end{equation}
which implies  rotational symmetry (the boundary $\mathcal{B}$ defines a solid of rotation). Since the covariant derivative is torsionless ($\nabla_{[a}\nabla_{b]}f=0$) and annihilates the three-dimensional Levi-Civita tensor $\varepsilon^{abc}$, the three-divergence $\nabla_an^a$ satisfies
\begin{equation}
\nabla_an^a = N'(\rho)\varepsilon^{abc}\partial_a\rho\partial_b\varphi\partial_c\eta=\frac{1}{\ell^3}\frac{N'(\rho)}{\sh \rho\ch\rho}.
\end{equation}
Given the ansatz \eref{ndef}, a solution to the boundary condition $\nabla_an^a=2/\ell$ is therefore given by
\begin{equation}
N(\rho) = \ell^2 \big(\sh^2\rho+ c\big).
\end{equation}
In the following, we will restrict ourselves to those configurations, where the integration constant $c$ is strictly positive, and we write, therefore
\begin{equation}
N(\rho) = \ell^2 \big(\sh^2\rho+ a^2\big),\label{Neq}
\end{equation}
for some constant $a> 0$. An additional constraint follows from the normalisation of the vector field $n^a$, which must be normalised to one, hence
\begin{align}
\nonumber
g_{ab}n^a n^b & = N^2(\rho)\,g^{ab}\,\partial_a\varphi\,\partial_b\varphi\, g^{cd}\,\partial_c\big(\eta-f(\rho)\big)\partial_d\big(\eta-f(\rho)\big)=\\
& = \frac{1}{\ell^4}N^2(\rho)\frac{1}{\sh^2(\rho)}\bigg[\frac{1}{\ch^2\rho}+\big(f'(\rho)
\big)^2\bigg]=1.\label{fdef1}
\end{align}
The function  $N(\rho)$ is already given in \eref{Neq}, and the normalisation of $n^a$ determines, therefore, a differential equation for $f(\rho)$, namely
\begin{equation}
\big[\tfrac{\di}{\di\rho}f(\rho)\big]^2 = \frac{\sh^2\rho}{[\sh^2\rho+a^2]^2}-\frac{1}{\ch^2\rho}.\label{fdef2}
\end{equation}
The left hand side must always be greater or equal to zero, which implies that the $\rho$-coordinate satisfies the inequality
\begin{equation}
\rho\geq \rho_o = \log\Big(\frac{1}{\sqrt{1-2a^2}}\Big).\label{rhoo}
\end{equation}

Next, we have to demonstrate that the resulting vector field $n^a$ defines a surface $\mathcal{B}\subset\H$ to which it lies orthogonal. Such a surface exists, if and only if the co-vector $n_a$ satisfies the Frobenius integrability condition,
\begin{equation}
\nabla_{[a}n_{b]} = \omega_{[a}n_{b]},
\end{equation}
for some one-form $\omega_a$. Going back to our ansatz \eref{ndef} for the vector field $n^a$, this is immediate to verify: the co-vector $n_a$ is given by
\begin{equation}
n_a = \frac{N(\rho) f'(\rho)}{\th\rho}\Big(\partial_a\eta+\frac{1}{\ch^2\rho}\frac{1}{f'(\rho)}\partial_a\rho\Big),
\end{equation}
and its exterior derivative $\di n$ satisfies, therefore, $\di n \propto \di\rho\wedge \di \eta\propto \di\rho\wedge n$, which implies, in turn, that the condition for the Frobenius integrability theorem is satisfied ($\omega\propto\di \rho$). The vector field $n^a$ is therefore indeed orthogonal to a two-dimensional submanifold $\mathcal{B}\subset\H$.

To understand how this surface $\mathcal{B}$ lies within $\H$, let us take $\rho$ and $\varphi$ as independent coordinates intrinsic to $\mathcal{B}$. If we then restrict ourselves to the negative  square root for $f'(\rho)$, i.e.\
\begin{equation}
\frac{\di f(\rho)}{\di\rho} = - \frac{\sqrt{(1-2 a^2)\sh^2\rho -a^4}}{(\sh^2\rho+a^2)\ch\rho},
\end{equation}
we find
\begin{equation}
\frac{\di\eta}{\di\rho}\bigg|_{\mathcal{B}} = \frac{1}{\ch\rho}\frac{\sh^2\rho+a^2}{\sqrt{(1-2 a^2)\sh^2\rho - a^4}},
\label{etarho}\end{equation}
which determines the dependence of the $\eta$-coordinate along $\mathcal{B}$. We will solve this differential equation implicitly below.\vspace{0.5em}

Having given a particular example for a hypersuface $\mathcal{B}$ that satisfies $K=2/\ell$, we now want to identify the corresponding holomorphic spinor field thereon. To this goal, let us first introduce complex coordinates that diagonalise the induced $\H$ line element on  $\mathcal{B}$, which is given by the pull-back
\begin{equation}
d\sigma^2:=\varphi^\ast_{\mathcal{B}} ds^2 = \ell^2 \sh^2\rho\Big(\frac{\ch^2\rho}{(1-2 a^2)\sh^2\rho-a^4}\di\rho^2+\di\varphi^2\Big).\label{sigmadef}
\end{equation}
We now look for a  complex coordinate 
\begin{equation}
z = \E^{x-\I\varphi},\label{zdef}
\end{equation}
that conformally maps the induced metric \eref{sigmadef} into the flat metric $q_{ab}=\partial_{(a}\,z\partial_{b)}\bar{z}$ on the punctured complex plane $\C-\{0\}$. In other words,
\begin{equation}
\di x = \frac{\di \sh\rho}{\sqrt{(1-2 a^2)\sh^2\rho-a ^4}}.
\end{equation}
Choosing initial conditions $\rho(x=0)=\rho_o$, with the minimal radius $\rho_o$ given in \eref{rhoo}, we infer the solution
\begin{equation}
\sqrt{1-2 a^2}\sh \rho = a^2 \ch\big(\sqrt{1-2 a^2}\,x\big).\label{rhodef1}
\end{equation}
Going back to \eref{etarho}, we can then also immediately infer $\eta$ as a function of $x$. With initial conditions $\eta(x=0)=0$, we find
\begin{equation}
\eta(x) = x - \frac{1}{2} {\log}{\bigg(\frac{1-a^2+\sqrt{1-2 a^2}\th(\sqrt{1-2 a^2}\,x)}{1-a^2-\sqrt{1-2 a^2}\th(\sqrt{1-2 a^2}\,x)}\bigg)}.\label{etax}
\end{equation}
For any fixed $a>0$ the functions $\eta(x)$ and $\rho(x)$ define, therefore, an embedding of the punctured complex plane into $\H$ such that the condition $K=2/\ell$ is satisfied. In the limit of $a\rightarrow1/\sqrt{2}$ we approach the asymptotic cylinder $\rho\rightarrow\infty$.

To determine the corresponding spinor field $\xi^A$ on $\mathcal{B}$, we now need to choose a cotriad that diagonalises the $\H$ metric \eref{gdef}. To cover the entire $\H$ space, we introduce the following rotating frame\footnote{If we introduce new coordinates $r:=\ell\rho$, and $x^1+\I x^2 = r\E^{\I\varphi}$, and $x^3:=\ell\eta$, the rotating frame $(\ref{e1frame}, \ref{e2frame}, \ref{e3frame})$ reduces in the Euclidean  $\ell\rightarrow\infty$ limit to the Cartesian frame $e^i=\di x^i$ in $\R^3$.}
\begin{subalign}
e^1 & = \ell(\cos \varphi\,\di\rho-\sh\rho\sin\varphi\,\di\varphi),\label{e1frame}\\
e^2 & = \ell(\sin \varphi\,\di\rho+\sh\rho\cos\varphi\,\di\varphi),\label{e2frame}\\
e^3 & = \ell\ch\rho\,\di\eta.\label{e3frame}
\end{subalign}
The components of the corresponding Levi-Civita spin connection are given by
\begin{subalign}
\omega^1 & = +\sh\rho\sin\varphi\,\di\eta,\\
\omega^2 & = -\sh\rho\cos\varphi\,\di\eta,\\
\omega^3 & = (\ch\rho-1)\,\di\varphi,
\end{subalign}
that satisfy the torsionless equation $\nabla e^i = \di e^i+\ou{\epsilon}{i}{lm}\omega^l\wedge e^m=0$, which determines $\omega^i$ as a function of the frame fields $e^i$. We can now proceed to identify the boundary spinor $\xi^A$. The defining property of $\xi^A$ is that it diagonalises the induced triad on $\mathcal{B}$,
\begin{equation}
\frac{1}{\sqrt{2}}\ou{\sigma}{A}{Bi}(\varphi^\ast_{\mathcal{B}}e^i)_a =-4\pi G\big(\xi^A\xi_Bm_a+\HC\big),
\end{equation}
where $\xi_A = \epsilon_{BA}\xi^B$ is the dual spinor. With respect to the Cartesian coordinates \eref{zdef}, the dyadic one-form $m_a\in T^\ast_\C\mathcal{B}$ is simply given by
\begin{equation}
m_a= \frac{1}{\sqrt{2}}\partial_a z.
\end{equation}
The boundary spinor $\xi^A$ can be inferred, therefore, immediately from the equation
\begin{align}\nonumber
\ou{\sigma}{A}{Bi}&\ou{e}{i}{a}\big(\partial^a_x+\I\partial^a_\varphi\big)=-8\pi G\,\xi^A\xi_B=\\
&=\frac{\ell}{\ch \rho}  
\begin{pmatrix}
a^2+\sh^2\rho & \big(a^2\sh(\sqrt{1-2 a^2}x+\sh\rho\,\ch\rho)\big)\E^{-\I\varphi}\\
\big(a^2\sh(\sqrt{1-2 a^2}x-\sh\rho\,\ch\rho)\big)\E^{-\I\varphi} & -(a^2+\sh^2\rho)
\end{pmatrix}.
\end{align}
Up to an overall undetermined sign, we thus find 
\begin{equation}
\xi^A=\Bigg(\begin{matrix}
\xi^0(x,\varphi)\\\xi^1(x,\varphi)
\end{matrix}\Bigg) = \sqrt{\frac{\ell}{8\pi G}}\frac{\I}{\sqrt{\ch\rho}}
\begin{pmatrix*}[l]
+\sqrt{\sh\rho\ch\rho+a^2\sh(\sqrt{1-2 a^2}\,x)}\,\E^{-\frac{x}{2}}\\
-\sqrt{\sh\rho\ch\rho-a^2\sh(\sqrt{1-2 a^2}\,x)}\,\E^{-\frac{x}{2}+\I\varphi}
\end{pmatrix*},
\end{equation}
where the $\rho$-coordiante has to be understood as an implicit function of $a$ and $x$ according to \eref{rhodef1}.
 
We are now left to demonstrate that $\xi^A$ defines a holomorphic spinor with respect to the $SL(2,\C)$ connection along the boundary. Since the connection is flat and the interior of the cylinder is simply connected, we can always find an $SL(2,\C)$ gauge element $g:\H\rightarrow SL(2,\C)$ such that the De\,Sitter connection
\begin{equation}
A_a = \frac{1}{2\I}\sigma_i\Big(\omega^i+\frac{\I}{\ell} e^i\Big)
\end{equation}
can be written as
\begin{equation}
A_a = g^{-1}\partial_a g.
\end{equation}
By integrating this equation along the $\eta$-axis and along the radial $\rho$-direction, we easily find 
\begin{equation}
g(\rho,\varphi,\eta) = \ch\left(\frac{\rho}{2}\right)
\begin{pmatrix}
\E^{-\frac{\eta}{2}} & 0\\
0 & \E^{+\frac{\eta}{2}}
\end{pmatrix}
+\sh\left(\frac{\rho}{2}\right)\begin{pmatrix}
0 & \E^{\frac{\eta}{2}-\I\varphi}\\
\E^{-\frac{\eta}{2}+\I\varphi} & 0
\end{pmatrix}\!\in SL(2,\C).
\end{equation}
Finally, we perform the parallel transport and map the boundary spinor $\xi^A(x,\varphi)$ back into the origin $(\rho=0,\eta=0)$ of the coordinate system. This is achieved by some straightforward matrix algebra and yields the holomorphic spinor\begin{equation}
\eta^A(z) := \ou{g}{A}{B}\big(\rho(x),\varphi,\eta(x)\big)\xi^B(x,\varphi) = \I a\sqrt{\frac{\ell}{8\pi G}}  
\Bigg(\begin{matrix}
1\\-z^{-1}\label{etaexampl}
\end{matrix}\Bigg).
\end{equation}
Hence the  spinor $\xi^A(x,\varphi)$ in the $\C^2$-spinor bundle over the boundary defines a holomorphic function $\eta^A(z)$ on the punctured complex plane $\C-\{0\}$. In the next section, we will demonstrate how to generalise this result to arbitrary genus 0 cylinders that are immersed\footnote{There may be a non-trivial winding that wraps the cylinder into itself. Such winding numbers play an important role in the evaluation of the non-perturbative spinfoam amplitude on a solid torus, see \cite{Dittrich:2018xuk}.} into Euclidean $\AdS$.
\subsection{Generic solution, monodromy, and deformed Gauss law}\label{sec3.2}
\noindent After having constructed an explicit solution\footnote{The boundary $\mathcal{B}$ is an example of Bryant's catenoid cousins, see \cite{AST_1987__154-155__321_0}.} of the bulk plus boundary field theory, we now need to understand the geometry of a generic such solution.

First of all, we introduce the two $SL(2,\C)$ connections,
\begin{subalign}
\ou{A}{A}{Ba} &= \frac{1}{2\I}\,\ou{\sigma}{A}{Bi}\,\big(\ou{\omega}{i}{a} + \frac{\I}{\ell}\ou{e}{i}{a}\big),\label{Adegf}\\
\ou{\bar{A}}{A'}{B'a} &= \frac{1}{2\I}\ou{\bar\sigma}{A'}{B'i}\big(\ou{\omega}{i}{a} - \frac{\I}{\ell}\ou{e}{i}{a}\big),\label{Abardef}
\end{subalign}
where primed spinor indices transform\footnote{At $\mathcal{B}$, generic such $SL(2,\C)$ transformations are no longer gauge directions on phase space, but become physical.} under the complex conjugate representation of $SL(2,\C)$. On shell, the Einstein and torsionless equations imply that the two connections are locally flat. Since the initial hyspersurface $\Sigma$ is assumed to be a genus zero disk, the general solution of the equations of motion can always be written in terms of a single-valued holonomy $g:\Sigma\rightarrow SL(2,\C)$,
\begin{equation}
\ou{A}{A}{Ba} =\ou{[g^{-1}\partial_a g]}{A}{B},\qquad \ou{A}{A'}{B'a} =\ou{[\bar{g}^{-1}\partial_a \bar{g}]}{A'}{B'}.\label{ADSconnectn}
\end{equation}
The dyadic one-forms $(m_a,\bar{m}_a)\in\Omega^1(\mathcal{B}:\C)$ are external background fields ($c$-numbers) on the covariant phase space. Their field variations vanish $\delta[m_a]=0$, and we can restrict ourselves, therefore, to the flat case
\begin{equation}
m_a = \frac{1}{\sqrt{2}}\partial_az,
\end{equation}
such that the $U(1)$ boundary spin connection vanishes, see \eref{Btors}. Since the De\,Sitter connections are flat, the boundary equations of motion, \eref{EOM2}, translate now into the ordinary Cauchy\,--\,Riemann differential equations,
\begin{equation}
\partial_{\bar{z}}\eta^A = 0,\label{EOM3}
\end{equation}
where we defined the parallel transported spinor
\begin{equation}
\eta^A = \ou{g}{A}{B}\xi^B.\label{etadef}
\end{equation}
If $\eta^A$ is single valued and has no singularities in $\mathcal{B}$, it admits the Laurent expansion,\footnote{A specific example for such spinor that describes a catenoid has been given in equation \eref{etaexampl} above.}
\begin{equation}
\eta^A(z) = \frac{1}{\sqrt{2\pi}}\sum_{n=-\infty}^\infty\eta^A_n z^n.\label{etaLaurent}
\end{equation}

 So far, we have solved, however, only one half of the boundary equations of motion, namely equation \eref{EOM2} that imposes $K=2/\ell$. In addition to the boundary condition on the extrinsic curvature, there are, however, also boundary conditions for the off-diagonal metric components, namely the \emph{gluing conditions} \eref{glucond}. In terms of the flat De\,Sitter connection $A_a=g^{-1}\partial_a g$, these gluing conditions translate now into the following constraint,\begin{equation}
\varphi^\ast_{\mathcal{B}}\ou{\big[g^{-1}\di g\big]}{A}{B}+\HC = -\frac{4\pi G}{\ell}\Big(\ou{[g^{-1}]}{A}{C}\eta^C\eta_D\ou{g}{D}{B}\di z + \HC\Big),\label{glucond2}
\end{equation}
where $\HC$ denotes the Hermitian conjugate with respect to the $SU(2)$ metric $\delta_{AA'}$, e.g.\ $\ou{[X^\dagger]}{A}{B}=\delta^{AB'}\ou{\bar{X}}{A'}{B'}\delta_{BA'}$. To disentangle the primed and unprimed spinor contributions to this equation, we consider the following ansatz for the $SL(2,\C)$ group element at the boundary,
\begin{equation}
g\big|_{\mathcal{B}} = hU,
\end{equation}
where $U\in SL(2,\C)$ is yet unspecified and $h$ is a holomorphic function $h:\C\rightarrow SL(2,\C)$ that satisfies the following holonomy equation
\begin{equation}
\frac{\di}{\di z}\ou{h}{A}{B} = - \frac{4\pi G}{\ell} \eta^A\eta_C \ou{h}{C}{B},\label{holdef}
\end{equation}
to some initial condition $h(z_o) = h_o \in SL(2,\C)$. If we insert this ansatz back into the gluing condition \eref{glucond2}, we immediately find that the function\footnote{The group element $g:\Sigma\rightarrow SL(2,\C)$ is single-valued, but the solutions $\ou{h}{A}{B}$ of \eref{holdef} may have a branch cut, which we can always put on the negative real axis, i.e.\ $\mathcal{B}' \simeq \C^\prime=\C - \R_{-}$.} $U:\C'\rightarrow SL(2,\C)$ must satisfy the following constraint equation,
\begin{equation}
[U^{-1}]\,\tensor[^2]{\di}{}U + \HC =0,
\end{equation}
where $\tensor[^2]{\di}{}$ is the exterior derivative on $\mathcal{B}$. The generic solution of this equation is $U(z,\bar{z}) = U_o \tilde{U}(z,\bar{z})$, where $U_o$ is a constant $SL(2,\C)$ element and $\tilde{U}(z,\bar{z})$ defines a map $\tilde{U}:\mathcal{B}'\rightarrow SU(2)$. Since the initial value $h_o$ of $h(z)$ is already arbitrary, we can assume without loss of generality $U_o = \bbvar{1}$ and $\tilde{U}(z,\bar{z})\in SU(2)$.

Since $g:\Sigma\rightarrow SL(2,\C)$ is single-valued (the disk $\Sigma$ has no handles or holes), there is one further and non-local constraint. For a general boundary spinor $\eta^A(z)$, such as the one that describes \emph{Bryant's catenoid cousins}, see \eref{etaexampl}, the solutions of the holonomy equation \eref{holdef} will have a branch cut that we can always put along the negative real axis, but $\ou{g}{A}{B}$ is single-valued, hence there is one additional constraint.
Suppose then that the initial point $z_o:|z_o|>0$ lies on the branch cut. If $\gamma(z_o\rightarrow z)$ is a family of paths $\gamma(z_o\rightarrow z):(0,1)\rightarrow\C-\R_-$ that connects\footnote{i.e.\ $\lim_{\varepsilon\searrow 0}\gamma(z_o\rightarrow z)(\varepsilon) = z_o$, $\lim_{\varepsilon\searrow 0}\gamma(z_o\rightarrow z)(1-\varepsilon) = z$, such that $\gamma(z_o\rightarrow z_o)$ denotes a closed loop that winds once around the origin $z=0$.} the fixed initial point $z_o$ with any other $z\in\C-\R_-$, the general solution of $h(z)$ is given by the path-ordered exponential,
\begin{equation}
h(z) = h(z,z_o)h_o\equiv\mathrm{Pexp}\Big(-\frac{4\pi G}{\ell}\smashoperator{\int_{\gamma(z_o\rightarrow z)}}\di z'\, \Omega(z')\Big)h_o,\label{propgtr}
\end{equation}
where $\Omega(z)$ defines the holomorphic and flat $SL(2,\C)$ connection,
\begin{equation}
\ou{\Omega}{A}{B}(z)\di z = \eta^A(z)\eta_B(z)\di z.\label{Omconnectn}
\end{equation}
If we wind once around the origin, a generic such holomorphic spinor\footnote{Such as the one that defines the hyperbolic catenoid cousins, see \eref{etaexampl}.} $\eta^A(z)$ will induce a non-trivial monodromy,
\begin{equation}
M = h_o^{-1}\mathrm{Pexp}\Big(-\frac{4\pi G}{\ell}\smashoperator{\oint_{\gamma(z_o\rightarrow z_o)}}\di z'\,\Omega(z)\Big)h_o\stackrel{\text{in general}}{\neq}\bbvar{1}.
\end{equation}
Since, however, $g:\Sigma\rightarrow SL(2,\C)$ is single-valued, and $g=hU$ is parametrised in terms of a holomorphic function $h(z)\in SL(2,\C)$ and an additional $SU(2)$ element $U(z,\bar{z})$, we must conclude that the monodromy lies in $SU(2)$, otherwise $g=hU$ cannot be single-valued. In other words,
\begin{equation}
M  = h_o^{-1}\mathrm{Pexp}\Big(-\frac{4\pi G}{\ell}\smashoperator{\oint_{\gamma(z_o\rightarrow z_o)}}\di z'\, \Omega(z)\Big)h_o \in SU(2).\label{Mdef}
\end{equation}
We must impose, therefore, an additional non-local closure constraint,
\begin{equation}
M^\dagger M = \bbvar{1}.\label{gauss}
\end{equation}
If this constraint is satisfied, we have a single-valued function $g(z,\bar{z})=h(z)U(z,\bar{z})$ on the boundary $\mathcal{B}$ of the cylinder.\footnote{To extend this function into the bulk, we write $g(z,\bar{z})$ as a product $g(z,\bar{z}) =  B(z,\bar{z})V(z,\bar{z})$, where $V$ is an $SU(2)$ element and $B$ is a Lorentz boost that can always be written as $B=\exp(X)$ for some $X^\dagger = X$. We now choose a new transversal and radial  coordinate $\rho\in(0,1]$ in $\mathcal{M}$, and define the function $g(\rho,z,\bar{z}):=\exp(\rho X(z,\bar{z}))V(z,\bar{z})$, which extends $g$ from the boundary into the interior. In these coordinates, the boundary $\mathcal{B}$ is the surface $\rho=1.$}

To understand the physical significance of the constraint \eref{gauss} on the monodromy, let us consider the $\ell\rightarrow \infty$ Euclidean limit of this equation \emph{of-shell}, i.e.\ prior to imposing the constraint \eref{gauss}. We define
\begin{equation}
\lim_{\ell\rightarrow\infty}\frac{M^\dagger(\ell) M(\ell)-\bbvar{1}}{\ell^{-1}} =: G,
\end{equation}
where the monodromy $M(\ell)$ depends  for given boundary conditions $h(z_o)=h_o$ and fixed connection coefficients $\ou{\Omega}{A}{B}(z)=\eta^A(z)\eta_B(z)$ implicitly on $\ell$. A short calculation gives,
\begin{equation}
\frac{\di}{\di \ell} M(\ell) = \frac{4\pi G}{\ell^2}\smashoperator{\oint_{\gamma(z_o\rightarrow z_o)}}\di z\, h_o^{-1}h(z_o,z) \Omega(z) h(z,z_o) h_o,
\end{equation}
where $h(z_o,z)$ is the parallel transport along the portion\footnote{N.B.\ $\gamma(z\rightarrow z_o)\circ\gamma(z_o\rightarrow z)=\gamma(z_o\rightarrow z_o)$.} $\gamma(z\rightarrow z_o)$ of the loop $\gamma(z_o\rightarrow z_o)$ that starts at $z$ and ends at $z_o$. In other words,
\begin{equation}
h(z_o,z)=\mathrm{Pexp}\Big(-\frac{4\pi G}{\ell}\smashoperator{\int_{\gamma(z\rightarrow z_o)}}\di z\,\Omega(z)\Big).
\end{equation}
Taking into account that $\lim_{\ell\rightarrow\infty}M(\ell) = \bbvar{1}$, we can use now D'Hopital's rule and find,
\begin{equation}
G=-4\pi G\Big[\smashoperator{\oint_{\gamma(z_o\rightarrow z_o)}}\di z\,h_o^{-1}\Omega(z) h_o +\HC\Big].\label{clos1}
\end{equation}
For $\ell\rightarrow\infty$, the $SL(2,\C)$ group element $g=h(z,z_o)h_o U(z,\bar{z})$ that defines the De\,Sitter connection turns now into $g=h_oU$, where $U$ is the holonomy of the $SU(2)$ spin connection, and $h_o$ is a constant $SL(2,\C)$ group element at the boundary $\mathcal{B}$. The integral \eref{clos1} is then nothing but the dressed integral\footnote{The $\mathfrak{su}(2)$-valued integrand is parallely transported into the frame over the reference point $z_o\in\mathcal{B}$.} of the triad, which is now parametrised in terms of spinors $\xi^A$ that define the flat triad $\xi^A\xi_Bm+\HC\propto \ou{\sigma}{A}{Bi}\varphi^\ast_{\mathcal{B}}e^i$ at the boundary. More specifically,
\begin{equation}
\lim_{L\rightarrow\infty}\frac{M^\dagger(\ell) M(\ell)-\bbvar{1}}{\ell^{-1}}  = \smashoperator{\oint_{\gamma(z_o\rightarrow z_o)}}\,U\sigma_i U^{-1} e^i=0.\label{flatclos}
\end{equation}
For three-dimensional gravity with a vanishing cosmological constant, this is nothing but the Gauss (or closure) constraint that generates rigid $SU(2)$ frame rotations on the phase space of the theory, see e.g.\ \cite{carlipbook, alexreview}. For a non-vanishing cosmological constraint $\Lambda= -\ell^{-2}$, the flat closure constraint \eref{flatclos} is replaced, therefore, by the group-valued constraint \eref{gauss} on the monodromy. We will see in the next section that this deformed closure constraint \eref{gauss} is still related to rigid $SU(2)$ frame rotations at the boundary. Similar deformed closure constraints have been identified in the discrete spinfoam approach to three-dimensional gravity, see for instance \cite{Dupuis:2014fya,Charles:2015lva,Bonzom:2014wva} and references therein.

\section{Covariant phase space, deformed Heisenberg algebra}\label{sec4}
\subsection{Covariant phase space and boundary energy momentum tensor}
\noindent The covariant pre-symplectic potential on a cross section $\Sigma$ of the cylinder $\mathcal{M}\simeq\Sigma\times$ can be now inferred from the first variation of the bulk plus boundary action \eref{actndef},
\begin{equation}
\delta S[e,\omega|\xi] = \mathrm{EOM}\cdot\delta + \Theta_{\Sigma^+}(\delta)-\Theta_{\Sigma^-}(\delta),
\end{equation}
where the equations of motion (EOM) include now \emph{both} the three-dimensional field equations in the bulk
\emph{and} the additional boundary field equations (\ref{EOM2}, \ref{glucond}). The reminder $\Theta_\Sigma(\delta)$ of the variation defines the pre-symplectic one-form on the covariant phase space, which is the space $\mathcal{P}_\Sigma$ of solutions of the bulk plus boundary field equations in a neighbourhood of $\Sigma$. On the cross section $\Sigma$, the pre-symplectic potential has then contributions from both the interior of the cylinder and its boundary,
\begin{align}
\Theta_\Sigma =  \frac{1}{8\pi G}\oint_{\Sigma}e_i\wedge\bbvar{d} \omega^i +\frac{\I}{\sqrt{2}}\oint_{\partial\Sigma}\Big[m
\,\xi_A\bbvar{d}\xi^A-\CC\Big],\label{thetadef1}
\end{align}
where $\bbvar{d}$ denotes the exterior derivative on the infinite-dimensional phase space $\mathcal{P}_\Sigma$. Poisson brackets are inferred from the covariant pre-symplectic two-form, which is given by the exterior functional derivative
\begin{equation}
\Omega_\Sigma = \bbvar{d}\Theta_\Sigma.\label{Omdef}
\end{equation}
For any two vector fields $\delta_1$ and $\delta_2$ on the covariant phase space $\mathcal{P}_\Sigma$, we then have,
\begin{equation}
\Omega_\Sigma(\delta_1,\delta_2) = \delta_1[\Theta_\Sigma(\delta_2)]- \delta_2[\Theta_\Sigma(\delta_1)] - \Theta_\Sigma\big([\delta_1,\delta_2]\big).
\end{equation}

To understand the gauge symmetries of the theory, we now have to identify the degenerate directions of the pre-symplectic two-form $\Omega_\Sigma$ on the covariant bulk plus boundary phase space $\mathcal{P}_\Sigma$. An internal $SU(2)$ frame rotation, which is connected to the identity, is generated by an infinitesimal gauge element $\Lambda^i:\mathcal{M}\rightarrow\mathfrak{su}(2)$. Such a gauge element can be lifted naturally into a vector field $\delta_{\Lambda}\in T\mathcal{P}_\Sigma$ on the covariant phase space,
\begin{subalign}
\delta_\Lambda[\omega^i] & = - \nabla\Lambda^i,\label{lambda1}\\
\delta_\Lambda[e^i] & = \ou{\epsilon}{i}{jk}\Lambda^je^k,\label{lambda2}\\
\delta_\Lambda[\xi^A] & = \ou{\tau}{A}{Bi}\Lambda^i\xi^B,\label{lambda3}
\end{subalign}
where $\nabla_a = \partial_a + [\omega_a,\cdot]$ is the covariant derivative of the $SU(2)$ connection $\ou{\omega}{i}{a}$. If $\delta\in T\mathcal{P}_\Sigma$ denotes now a second linearly independent vector field on the covariant phase space, i.e.\ a linearised solution of the bulk plus boundary field equations, we easily find
\begin{align}
\nonumber \Omega_\Sigma(\delta_\Lambda,\delta) & = \frac{1}{8\pi G}\int_\Sigma\Big[\epsilon_{ijk}\Lambda^j e^k\wedge \delta\omega^i+\delta e_i\wedge\nabla\Lambda^i\Big]+\frac{1}{\sqrt{2}}\oint_{\partial\Sigma}\Big[m\sigma_{ABi}\Lambda^i\xi^B\delta\xi^A+\CC\Big]=\\
& = -\frac{1}{8\pi G}\int_\Sigma \delta[\nabla e_i]\wedge\Lambda^i-\frac{1}{8\pi G}\oint_{\partial\Sigma}\delta\Big[e_i-\frac{4\pi G}{\sqrt{2}}\big(m\xi_A\xi_B-\bar{m}\xi^\dagger_A\xi^\dagger_B\big)\ou{\sigma}{AB}{i}\Big]\Lambda^i=0.
\end{align}
The first term in the second line vanishes thanks to the torsionless condition \eref{TEOM}, and the second term vanishes  thanks to the gluing conditions \eref{glucond} that allow us to parametrise the pull back of the triad in terms of the boundary spinor $\xi^A$. Simulataneous frame rotations (\ref{lambda1}, \ref{lambda2}, \ref{lambda3}) of the bulk plus boundary fields define, therefore, an exact gauge symmetry of the coupled bulk plus boundary system.

Next, we consider the action of a diffeomorphism $\varphi =\exp(V)$, which is generated by a vector field $V^a\in T\mathcal{M}$, on the covariant phase space. A generic such diffeomorphism will violate our conformal boundary conditions \eref{bndrycond}. It is easy to see, however, that a vector field $V^a$, whose restriction to the boundary defines a conformal Killing vector of the boundary metric $q_{ab} = 2 m_{(a}\bar{m}_{b)}$ defines a symmetry of the bulk plus boundary field theory. Consider, therefore,  such a vector field,
\begin{equation}
V_t^a\in T\mathcal{M}:V_t^a\big|_{\mathcal{B}} = t^a\in T\mathcal{B}: D_{(a} t_{b)} = \frac{1}{2}D_ct^c q_{ab},
\end{equation}
where $D_a$ is the boundary covariant derivative that annihilates $q_{ab}$ as well as the dyadic frame fields $(m_a,\bar{m}_a)$, see (\ref{Btors}, \ref{Dxi2}). In addition, tensor indices at the boundary are raised and lowered with respect to the fiducial boundary metric $q_{ab}$ and its inverse, i.e.\ $t_a = q_{ab}t^b\in T^\ast\mathcal{B}$. Any such vector field $V_t$, can be then lifted into a vector field $\delta_t$ on the covariant phase space. Its components are given by
\begin{subalign}
\delta_t[\omega^i] & = V_t\hook F^i,\label{tdiffeo1}\\
\delta_t[e^i] &= V_t\hook (\nabla e^i)+\nabla(V_t\hook e^i)=\nabla V^i_t,\label{tdiffeo2}\\
\delta_t[\xi^A] &= t^aD_a\xi^A + \frac{1}{2}\bar{m}^aD_aN\xi^A,\label{tdiffeo3}
\end{subalign}
where $N=t^am_a$ is the holomorphic component of the conformal Killing vector, i.e.\ $m^aD_a N=0$. Consider then the following boundary integral
\begin{equation}
H[N] = \frac{\I}{\sqrt{2}}\oint_{\partial\Sigma}\Big[N \xi_AD\xi^A-\frac{2\sqrt{2}\,\pi G}{\ell}
N\bar{m}\|\xi\|^4-\CC\Big],
\end{equation}
where $\|\xi\|^2=\delta_{AA'}\xi^A\bar{\xi}^{A'}$ denotes the $SU(2)$ norm of the spinor $\xi^A$.
To demonstrate that $\delta_t$ is the Hamiltonian vector field of $H[N]$, we compute the variation of $H[N]$ on the covariant phase space,
\begin{align}\nonumber
\delta\big[H[N]\big] &=\frac{\I}{\sqrt{2}}\oint_{\partial\Sigma}\Big[2N\delta[\xi_A]D\xi^A+DN\delta[\xi_A]\xi^A-\frac{2\sqrt{2}\pi G}{\ell}N\bar{m}\|\xi\|^2\delta[\|\xi\|^2]+\\\nonumber
&\hspace{5em}-\frac{1}{2\I}N\xi_A\xi_B\ou{\sigma}{AB}{i}\delta[\omega^i]-\CC\Big]=\\\nonumber
&=\frac{\I}{\sqrt{2}}\oint_{\partial\Sigma}\Big[2m\delta[\xi_A]\delta_t[\xi^A]+2N\bar{m}\delta[\xi_A]m^aD_a\xi^A-2m\bar{N}\delta[\xi_A]m^aD_a\xi^A+\\
&\hspace{5em}-\frac{1}{2\I}N\xi_A\xi_B\ou{\sigma}{AB}{i}\delta[\omega^i]-\frac{4\sqrt{2}\pi G}{\ell}N\bar{m}\|\xi\|^2\delta[\|\xi\|^2]-\CC\Big].
\end{align}
We can now also use the boundary equations of motion \eref{EOM2}, which imply
\begin{equation}
m^a D_a\xi^A = \frac{2\sqrt{2}\pi G}{\ell}\|\xi\|^2\delta^{AA'}\bar{\xi}_{A'}.
\end{equation}
And we are therefore left with the expression,
\begin{equation}
\delta\big[H[N]\big] = \frac{\I}{\sqrt{2}}\oint_{\partial\Sigma}\Big[2m\,\delta[\xi_A]\delta_t[\xi^A]-\frac{1}{2\I}N\xi_A\xi_B\ou{\sigma}{AB}{i}\delta[\omega^i]-\CC\Big].\label{deltaH}
\end{equation}
We now want to demonstrate that $\delta_t$ is integrable and that it is indeed generated by the Hamiltonian $H[N]$. We thus pick a second linearly independent tangent vector on the covariant phase space and contract both vector fields with the pre-symplectic two-form. We then have,
\begin{align}\nonumber
\Omega_\Sigma(\delta_t,\delta) &= \frac{1}{8\pi G}\int_\Sigma\Big[\nabla V^i_t\wedge \delta[\omega_i]-\delta[e_i]\wedge V_t\hook F^i\Big]+\frac{\I}{\sqrt{2}}\oint_{\partial\Sigma}\Big[2m\delta_t[\xi_A]\delta[\xi^A]-\CC\Big]=\\\nonumber
&=\frac{1}{8\pi G}\int_\Sigma\Big[- (V_t\hook e^i)\delta F_i-\delta[e_i]\wedge V_t\hook F^i\Big]+\\\nonumber
&\hspace{5em}+\oint_{\partial\Sigma}\Big[\frac{1}{8\pi G} (t\hook e_i)\delta[\omega^i]+\frac{\I}{\sqrt{2}}\big(2m\delta_t[\xi_A]\delta[\xi^A]-\CC\big)\Big].
\end{align}
The Einstein equations \eref{FEOM} imply that the first  term vanishes.
%\begin{equation}
%(V_t\hook e^i)\delta F_i+\delta[e_i]\wedge V_t\hook F^i=-\Lambda\epsilon_{ijk}V_t^i\delta[e^j]\wedge e^k+\Lambda\epsilon_{ijk}\delta[e^i]\wedge e^jV_t^k=0.
%\end{equation}
 The second term, on the other hand, can be written in terms of the boundary spinors alone: going back to the gluing conditions \eref{glucond}, and comparing the resulting expression with $\delta[H[N]]$, we find
\begin{equation}
\Omega_\Sigma(\delta_t,\delta)  = \frac{\I}{\sqrt{2}}\oint_{\partial\Sigma}\Big[\frac{1}{2\I} N\xi_A\xi_B\ou{\sigma}{AB}{i}\delta[\omega^i]-2m\delta[\xi_A]\delta_t[\xi^A]-\CC\Big]=-\delta\big[H[N]\big].
\end{equation}
We have thus integrated the Hamiltonian field equations for any bulk diffeomorphism that is generated by a vector field $V^a_t\in T\mathcal{M}$, whose restriction to the boundary $\mathcal{B}$ defines a conformal Killing vector $t^a = V^a_t|_{\mathcal{B}}\in T\mathcal{B}$ that preserves the conformal structure at the boundary, i.e.\ $\mathcal{L}_t q_{ab}\propto q_{ab}$. There is a further simplification that will prove very useful in the following: if we reintroduce the $SL(2,\C)\times U(1)$ boundary covariant derivative $\mathcal{D}_a$, as defined in \eref{AdSD}, we can write
\begin{equation}
H[N]=\frac{\I}{\sqrt{2}}\oint_{\partial\Sigma}\big[N\xi_A\mathcal{D}\xi^A-\CC\big]=\oint_{\partial\Sigma} dv^a T_{ab} t^b,\label{HTdef}
\end{equation}
where $N = t^a{m}_a$ denotes the holomorphic component of the conformal Killing vector, $dv^a = \I\bar{m}^a m+\CC \in T\mathcal{B}\otimes T^\ast\mathcal{B}$ is the vector-valued line element, and $T_{ab}$ denotes the Brown\,--\,York boundary quasi-local stress-energy tensor \cite{BrownYork, Szabados:2004vb},
\begin{equation}
T_{ab}=\frac{1}{8\pi G}\Big(K_{ab}-\frac{1}{\ell}h_{ab}\Big).\label{Tdef}
\end{equation}
Notice that the energy momentum tensor is traceless, because the conformal boundary conditions imply $K=2/\ell$, see \eref{bndrycond2}. Therefore, $T_{ab}$ is completely specified by its holomorphic component, which determines the shear $\sigma=8\pi G\bar{m}^a\bar{m}^bT_{ab}$ of the boundary $\mathcal{B}$. %The isomorphism between spinors and vectors allows us to write the shear $\sigma$ as a specific holomorphic spin coefficient,
%\begin{align}\nonumber
%\sigma:=&\bar{m}^a\bar{m}^bT_{ab} = \bar{m}^a\bar{m}^b\nabla_an_b =\\
%=&-\frac{8\pi G}{\sqrt{2}}\xi^A\xi^B\bar{m}^aD_a\Big(\frac{\xi^{\phantom{\dagger}}_{(A}\xi^\dagger_{B)}}{\|\xi\|^2}\Big)=\frac{8\pi G}{\sqrt{2}}\xi_A\bar{m}^aD_a\xi^A=\frac{8\pi G}{\sqrt{2}}\xi_A\bar{m}^a\mathcal{D}_a\xi^A,\label{sigmaxi}
%\end{align}
%where we used the $SL(2,\C)\times U(1)$ boundary connection \eref{AdSD} to arrive at the last equality. The boundary field equations \eref{EOM2} impose that $\xi^A$ is holomorphic, which implies, in turn, that the shear $\bar{\sigma}$ is holomorphic, i.e.\ $m^aD_a\sigma=0$, in addition to the covariant conservation of $T_{ab}$, i.e.\ $D_a\ou{T}{a}{b}=0$.

\subsection{Extended phase space, Dirac bracket, deformed Heisenberg algebra}
\noindent The purpose of this section is to establish the Poisson commutation relations between the fundamental boundary modes on the physical phase space. Our starting point will be the parametrisation of a generic solution of the bulk plus boundary field equations in terms of the mode expansion \eref{etaLaurent}. Given this parametrisation, we will then compute the pull-back of the pre-symplectic two-form \eref{Omdef} with respect to the sequence of maps $\eta^A_n\rightarrow \eta^A(z)=\frac{1}{\sqrt{2\pi}}\sum_n\eta^A_nz^n\rightarrow \xi^A = \ou{[g^{-1}]}{A}{B}\eta^B$, which is induced by the Laurent expansion \eref{etaLaurent} of the boundary spinor $\xi^A$.% This procedure will allow us to infer the Poisson commutation relations for components $\eta^A_n$ of the mode expansion. 

In the last section, we have identified two contributions to the pre-symplectic potential, namely a boundary term $\propto \xi_A\bbvar{d}\xi^A$ for the gravitational edge modes and a contribution $\propto e_i\wedge\bbvar{d}\omega^i$ coming from the bulk. Let us consider the bulk integral first. The general solution of the field equations (\ref{TEOM}, \ref{FEOM}) in the interior is given by a flat $SL(2,\C)$ connection,
\begin{equation}
A_a = g^{-1}\partial_a g = \frac{1}{2\I}\sigma_i\Big(\ou{\omega}{i}{a}+\frac{\I}{\ell}\ou{e}{i}{a}\Big).
\end{equation}
If we now insert this parametrisation back into the pre-symplectic two-form \eref{Omdef}, we immediately recover the pre-symplectic two-form for three-dimensional gravity in the familiar Chern\,--\,Simons formulation of three-dimensional gravity,\footnote{The symbol ``${\reflectbox{\rotatebox[origin=c]{180}{\fontsize{8pt}{8pt}$\bbvar{V}$}}}$'' combines the wedge product on the infinite-dimensional phase space with the ordinary wedge product on spacetime: if $\delta_1$ and $\delta_2$  are vector fields on phase space, and $\alpha$ and $\beta$ are $p$-form fields on spacetime, $(\bbvar{d}\alpha\,{\reflectbox{\rotatebox[origin=c]{180}{\fontsize{8pt}{8pt}$\bbvar{V}$}}}\,\bbvar{d}\beta)(\delta_1,\delta_2):=\delta_1[\alpha]\wedge\delta_2[\beta]-\delta_2[\alpha]\wedge\delta_1[\beta]$.}
\begin{equation}
\int_\Sigma \bbvar{d}e_i\bbwedge\bbvar{d}\omega^i = \frac{\I \ell}{2}\int_{\Sigma}\operatorname{Tr}\Big(\bbvar{d}\big(g^{-1}\di g\big) \bbwedge\bbvar{d}\big(g^{-1}\di g\big)\Big)+\CC,\label{Om2form1}
\end{equation}
see \cite{Carlip:2005zn,Witten:1988hc}. The functional differential of the connection satisfies $\bbvar{d} A =\bbvar{d}(g^{-1}\di g)= g^{-1}\di (\bbvar{d}g g^{-1})g$, which implies that the bulk integral \eref{Om2form1} collapses into a total exterior derivative. We are now left with the boundary integral
\begin{equation}
\Omega_\Sigma = \I\oint_{\partial\Sigma}\bigg[\frac{\ell}{16\pi G}\operatorname{Tr}\left(g^{-1}\bbvar{d}g\bbwedge \bbvar{d}\big(g^{-1}\di g\big)\right)+\frac{m}{\sqrt{2}}\bbvar{d}\xi_A\bbwedge\bbvar{d}\xi^A-\CC\bigg].\label{Omdef2}
\end{equation}
On the covariant phase space, the boundary fields $\xi^A$ and $g\big|_{\mathcal{B}}$ are not completely independent, because there are  boundary equations of motion that introduce a coupling between the boundary fields. There is the holomorphicity condition \eref{EOM2} for the boundary spinor $\xi^A$, but there are also the gluing conditions
\begin{equation}
\varphi^\ast_{\mathcal{B}}\tensor{\big(g^{-1}\di g\big)}{^{A}_{B}} + \HC = -\frac{8\pi G}{\ell}\frac{1}{\sqrt{2}}\big(\xi^A\xi_B m+\HC\big),\label{glucond2}
\end{equation}
where the Hermitian conjugate is taken with respect to the $SU(2)$ Hermitian metric $\delta_{AA'}$. To impose these constraints, we proceed now as in \hyperref[sec3.2]{section 3.2} above. First of all, we note that the dyadic one-forms $(m_a,\bar{m})\in T^\ast_\C\mathcal{B}$ are a background field on phase space, hence $\bbvar{d}m =0$. Working on a fixed Riemann surface $\mathcal{B}=\C-\{0\}$, which has the topology of an infinite cylinder with open ends, we can now choose  Cartesian coordinates $z : m=\frac{1}{\sqrt{2}}\di z$ such that the fiducial boundary metric $q_{ab}= 2m_{(a}\bar{m}_{b)}$ is diagonal. Given these coordinates, the map $g:\mathcal{B}\rightarrow SL(2,\C)$ splits now into a holomorphic part $h:\mathcal{B}\rightarrow SL(2,\C)$ and a function $U:\mathcal{B}\rightarrow SU(2)$ that takes values in $SU(2)$, such that $g=hU$ is single-valued. Given this parametization of the boundary fields, the holomorphicity condition \eref{EOM2} for the boundary spinor $\xi^A$, i.e.\  $m^a\mathcal{D}_a\xi^A=0$, turns now into the ordinary Cauchy\,--\,Riemann differential equations $\partial_{\bar{z}}\eta^A=0$ for $\eta^A(z)$, where $\xi^A$ is related to $\eta^A$ via $\xi^A = \ou{[U^{-1}h^{-1}]}{A}{B}\eta^B$.

If we now want to use this parametrisation at the level of the pre-symplectic two-form \eref{Omdef2}, we have to take into account that the gluing conditions \eref{glucond2} translate into a \emph{constraint} between $\eta\otimes\eta$ and $h^{-1}\di h$, namely 
\begin{equation}
\ou{J}{A}{B}[\eta,h](z) = 0,\label{Jcons}
\end{equation}
where we defined the following functional on the extended phase space of field configurations $\ou{h}{A}{B}(z)$ and $\eta^{A}(z)$,
\begin{equation}
\ou{J}{A}{B}[\eta,h](z) = \frac{\I}{2}\Big[\eta^A(z)\eta_B(z)+\frac{\ell}{4\pi G}\tensor{\big[\partial_zh(z)h^{-1}(z)\big]}{^A_B}\Big]\in \mathfrak{sl}(2,\C).\label{Jdef}
\end{equation}
In addition to $\ou{J}{A}{B}(z)=0$, there is one further non-local constraint: the boundary fields $\ou{g}{A}{B}$ and $\xi^A$ are single-valued, but $h(z)$ may pick up a monodromy around the origin $z=0$.\footnote{This happens already for Bryant's curved catenoids, where $\eta^A(z)$ has a pole at the origin, see \eref{etaexampl}.} In fact, the group element $\ou{g}{A}{B}(z)$ is single-valued, if and only if the following additional non-local and complex-valued constraints are satisfied
\begin{equation}
C = h(z_o^+)U(z_o^+) - h(z_o^-) U(z_o^-)=0,\label{Cdef}
\end{equation}
where we have put the branch cut along the negative real axis.\footnote{The boundary points $z_o^\pm$ lie above and below the branch cut, $U(z_o^\pm)=\lim_{\varepsilon\searrow 0} U(z_o\pm\I\varepsilon)$.} As explained in \hyperref[sec3.2]{section 3.2} above, the constraints $C=0$ can be seen as a deformed version of the closure constraint $\oint_{\partial\Sigma} U^{-1}\sigma_iUe^i=0$ for the triadic fluxes in the flat $\Lambda\rightarrow 0$ limit \cite{carlipbook, alexreview}.

If we now insert this parametrisation into the pre-symplectic two-form \eref{Omdef2}, we find after some straightforward algebra that
\begin{align}\nonumber
\Omega_\Sigma^\uparrow & = \frac{\I}{2}\oint_{\mathcal{C}}\bigg[\di z\,\bbvar{d}\eta_A\bbwedge\bbvar{d}\eta^A-\frac{\ell}{8\pi G}\operatorname{Tr}\left(h^{-1}\bbvar{d} h\bbwedge \di\big(h^{-1}\bbvar{d}h\big)\right)-\CC\bigg]+\\
&\hspace{15em}+\bigg[\frac{\I \ell}{16\pi G}\operatorname{Tr}\left(h^{-1}\bbvar{d}h\bbwedge \bbvar{d}U U^{-1}\right)\Big|_{\partial{\mathcal{C}}} +\CC\bigg].\label{Omdef3}
\end{align}
Compared to the symplectic two-form in the $g$-$\xi$-representation \eref{Omdef2}, there is now an additional boundary term appearing. The geometric origin of this boundary contribution has to do with the  monodromy \eref{Mdef} around the origin: the boundary fields $\eta^A(z)$ and $\partial_z hh^{-1}$ are single-valued in $\C-\{0\}$, but the $SL(2,\C)$ group element $\ou{h}{A}{B}(z)$, which is the path-ordered exponential of $\eta\otimes\eta$, see\eref{propgtr}, may have a non-trivial $SU(2)$ monodromy. Accordingly, we introduce a branch cut along the negative real axis, such that the contour $\mathcal{C}$ defines a path in the complex plane that starts at some point $z_o^-$ on the negative real axis and winds once around the origin.\footnote{The integral $\oint_{\mathcal{C}}\di f = f\big|_{\mathcal{C}}$ denotes the difference $\lim_{\varepsilon\searrow 0} \big(f(z_o+\I\varepsilon)-f(z_o-\I\varepsilon)\big)\equiv f(z_o^+)-f(z_o^-).$}

The constraint \eref{Cdef} is related to residual and global $SU(2)$ gauge transformations. This can be seen as follows: consider the following vector field on phase space, which acts as a left-invariant derivative on the $SU(2)$ coordinates
\begin{equation}
Y_i[U(z_o^\pm)]=-U(z_o^\pm)\tau_i,\label{Yvec}
\end{equation}
but vanishes otherwise,
\begin{equation}
Y_i[h(z)]=0,\qquad Y_i[\bar{h}(z)]=0,\qquad Y_i[\eta^A(z)]=0,\qquad Y_i[\bar{\eta}^{A'}(z)]=0.
\end{equation}

Consider then a field variation $\delta$ that lies tangential to the $C=0$ constraint hypersurface, i.e.\ $\delta[C]=0$ with $C=0$ denoting the constraint \eref{Cdef} on the monodromy. The vector field $Y_i$ defines a degenerate direction of $\Omega_\Sigma$ and it defines, therefore, a gauge symmetry,
\begin{align}\nonumber
\Omega_\Sigma^\uparrow(Y_i,\delta)&=\frac{\I \ell}{16\pi G}\Big[\mathrm{Tr}\big(\tau_i U^{-1}(h^{-1}\delta h)U\big)-\CC\Big]\Big|_{\partial\mathcal{C}}=\\
&=\frac{\I \ell}{16\pi G}\Big[\mathrm{Tr}\big(\tau_i (U^{-1}h^{-1}\delta[hU])\big)-\CC\Big]\Big|_{\partial\mathcal{C}}=0.
\end{align}

To compute the Poisson brackets between the Laurent modes $\eta^A_n$, we consider now an extended phase space $\mathcal{P}_\Sigma^\uparrow$, whose coordinates are given by the $SU(2)$ elements $U(z_o^\pm)$ at the marked boundary points $\partial\mathcal{C}=\{z^+_o\}\cup\{z^-_o\}$, by the field configurations of $\eta^A(z)$, which is holomorphic in $\C-\{0\}$, and by $\ou{h}{A}{B}(z)$, which has a branch cut along the negative real axis, while the corresponding Maurer\,--\,Cartan form $\partial_z hh^{-1}$ is holomorphic in $\C-\{0\}$. The Poisson brackets $\{\cdot,\cdot\}^\uparrow$ on this extended phase space $\mathcal{P}_\Sigma^\uparrow$ are determined then by the symplectic two-form \eref{Omdef3}. To recover the Poisson commutation relations on the physical phase space, we have to impose the constraints (\ref{Jdef}, \ref{Cdef}) and perform the symplectic reduction.

To impose the constraints, it is convenient to introduce $\mathfrak{sl}(2,\C)$-valued smearing functions $\ou{f}{A}{B}(z)$ that are continuous across the branch cut  along the negative real axis. We are thus defining the smeared Kac\,--\,Moody generators
\begin{equation}
J[f] := \oint_{\mathcal{C}}\di z\,\ou{f}{A}{B}(z)\ou{J}{B}{A}[\eta,h](z)\equiv\oint_{\mathcal{C}}\di z\,\operatorname{Tr}(fJ).
\end{equation}
Using $\bbvar{d}(\partial_z h h^{-1}) = h\partial_z(h^{-1}\bbvar{d} h)h^{-1}$, we compute the functional differential of $J[f]$,
\begin{equation}
\bbvar{d}J[f] = - \I\oint_{\mathcal{C}} \left[\di z\, f^{AB}\eta_A\bbvar{d}\eta_B-\frac{\ell}{8\pi G}\mathrm{Tr}\big(h^{-1}fh\,\di(h^{-1}\bbvar{d} h)\big)\right].\label{diffJ}
\end{equation}
The constraints define an $SL(2,\C)$ Kac\,--\,Moody algebra. There is a central charge and the constraints are second-class. This can be seen as follows: consider first the following complexified vector field $X_f\in \big(T\mathcal{P}_\Sigma^\uparrow\big)_\C$, whose components on phase space are given by
\begin{subalign}
\ou{\big[h^{-1}(z)X_f[h(z)]\big]}{A}{B} &=\ou{f}{A}{B}(z),\label{Jactn1}\\
X_f[\eta^A(z)]&=\ou{f}{A}{B}(z)\eta^B(z).\label{Jactn2}
\end{subalign}
All other components vanish:
\begin{equation}
\bar{h}^{-1}X_f[\bar{h}] =0,\qquad X_f[\bar{\eta}^{A'}(z)] =0, \qquad X_f[U(z_o^\pm)]=0.
\end{equation}
We now want to demonstrate that $X_f$ is the Hamiltonian vector field of $J[f]$ provided the closure constraint \eref{Cdef} is satisfied. We proceed as in above: consider a second linearly independent field variation $\delta$ on the extended phase space, and contract both vector fields with $\Omega_\Sigma^\uparrow$. We obtain
\begin{equation}
\Omega^\uparrow_\Sigma(X_f,\delta) = -\delta J[f]+\frac{\I \ell}{16\pi G}\Big[\mathrm{Tr}\big(f\delta[g]g^{-1}\big)\Big|_{\partial\mathcal{C}}-\CC\Big],
\end{equation}
where $g=hU$. On the constraint hypersurface, where the closure constraint  \eref{Cdef} is satisfied, the map $g:\C-\{0\}\rightarrow SL(2,\C)$ is continuous across the branch cut and the last boundary term disappears. Up to terms constrained to vanish, the Hamiltonian vector field of $J[f]$ is given, therefore, by $X_f$. This in turn implies that we can immediately infer the constraint algebra,\footnote{Notice that the vector field $X_f$ preserves the constraint \eref{Cdef}, hence $X_i[C]=0$.}
\begin{align}
\nonumber
\big\{J[f],J[f']\big\}^\uparrow&\approx X_f\big[J[f']\big]=\I\oint_{\mathcal{C}}\bigg[\di z\, \ou{[f']}{A}{C}\ou{f}{C}{B}\eta_A\eta^B+\frac{\ell}{8\pi G}\mathrm{Tr}\big(h^{-1}f'h\,\di(h^{-1}fh)\big)\bigg]=\\
\nonumber&=\frac{\I}{2}\oint_{\mathcal{C}}\bigg[\ou{\big[f',f\big]}{A}{B}\eta^B\eta_A+\frac{\ell}{4\pi G}\mathrm{Tr}\Big(\big[f',f\big]\di hh^{-1}\Big)\bigg]+\frac{\I \ell}{8\pi G}\oint_{\mathcal{C}}\mathrm{Tr}\big(f'\di f\big)=\\
&=-J\big[[f,f']\big]+\frac{\I \ell}{8\pi G}\oint_{\mathcal{C}}\mathrm{Tr}\big(f'\di f\big),\label{JJpoiss}
\end{align}
where $\approx$ denotes equality up to terms that vanish on the $C=0$ constraint hypersurface and $[\cdot,\cdot]$ is the $\mathfrak{sl}(2,\C)$ Lie bracket in the fundamental spin $(\tfrac{1}{2},0)$ representation,
\begin{equation}
\tensor{[f,f']}{^A_B} = \ou{f}{A}{C}\ou{[f']}{C}{B} - \ou{[f']}{A}{C}\ou{f}{C}{B},\qquad [f,f']^{(AB)} = 2\ou{f}{(A}{C}[f']^{B)C}.
\end{equation}
The $SL(2,\C)$ Kac\,--\,Moody algebra \eref{JJpoiss} is anomalous. There is a central charge and the constraint algebra is, therefore, {second-class}. To infer the Poisson commutation relations on the physical phase space, we introduce the Dirac bracket. In this context, it is now useful to introduce the mode expansion,
\begin{equation}
J^i_n := J[\tau^i z^{-n}].\label{Jmode}
\end{equation}
From \eref{JJpoiss}, we can then immediately infer the Poisson commutation relations for the Kac\,--\,Moody constraints,
\begin{subalign}
\big\{J^i_n, J^k_m\big\}^\uparrow &\approx -\ou{\epsilon}{ik}{l}J^l_{n+m} - \frac{\ell}{8 G}n \delta_{n+m}\delta^{ik},\label{JJ1}\\
\big\{\bar{J}^i_n, \bar{J}^k_m\big\}^\uparrow &\approx -\ou{\epsilon}{ik}{l}\bar{J}^l_{n+m} - \frac{\ell}{8 G}n \delta_{n+m}\delta^{ik},\label{JJ2}
\end{subalign}
where $\delta_n=1$ if $n=0$, and $\delta_n=0$ otherwise. The zero mode $J^i_{n=0}$ is first-class (the constraint $\lambda_i J^i_{n=0}+\bar{\lambda}_i J^i_{n=0}$ generates global $SL(2,\C)$ frame rotations), all other constraints are second-class. On the physical phase space, where all constraints are satisfied, the Poisson brackets are given now by the Dirac bracket $\{\cdot,\cdot\}$, which is obtained by removing the unphysical $J$-directions from the auxiliary Poisson brackets $\{\cdot,\cdot\}^\uparrow$ on the extended phase space. In other words, 
\begin{equation}
\big\{F,G\big\} = \big\{F,G\big\}^\uparrow - \frac{8G}{\ell}\sum_{n\neq 0}\frac{1}{n}\big\{F,J^i_n\big\}^\uparrow\delta_{ik}\big\{J^k_{-n},G\big\}^\uparrow- \frac{8G}{\ell}\sum_{n\neq 0}\frac{1}{n}\big\{F,\bar{J}^i_n\big\}^\uparrow\delta_{ik}\big\{\bar{J}^k_{-n},G\big\}^\uparrow,\label{Dbrack}
\end{equation}
where $F=F[h,\eta,U]$ and $G\equiv[h,\eta,U]$ are functionals on the extended phase space $\mathcal{P}_\Sigma^\uparrow$.

On the extended phase space, which is equipped with the symplectic two-form $\Omega^\uparrow_\Sigma$, the Laurent modes $\eta^A_n$ of $\eta^A(z)$  generate an infinite-dimensional Heisenberg algebra,\footnote{If we introduce for any $n\geq 0$ the position and momentum modes $q^A_n=\eta^A_n$ resp.\ $p_A^n=\epsilon_{BA}\eta^B_{-n-1}$, we recover the usual canonical commutation relations $\{p_A^n,q^B_m\}=-\delta^B_A\delta^n_m$.}
\begin{equation}
\big\{{\eta}^A_n,{\eta}^B_m\big\}^\uparrow = -\epsilon^{AB}\delta_{m+n+1},\qquad \big\{\bar{\eta}^{A'}_n,\bar{\eta}^{B'}_m\big\}^\uparrow = -\bar{\epsilon}^{A'B'}\delta_{m+n+1}.\label{Hpoiss}
\end{equation}
The action of the Kac\,--\,Moody constraints $J^i_n$ on the Fourier modes $\eta^A_n$ is immediate: $\{J^i_n,\eta^A_m\}^\uparrow=\tensor{\tau}{^A_B^i}\,\eta^B_{n+m}$. In other words, the second-class constraints $J_n^i$ do not commute with $\eta^A_n$, and the commutation relations for the Fourier modes $\eta^A_n$ will be significantly changed by the introduction of the Dirac bracket \eref{Dbrack}.
In fact, we find the following deformation of the Heisenberg algebra \eref{Hpoiss} for the boundary modes on the physical phase space,
\begin{subalign}
\big\{{\eta}^A_n,{\eta}^B_m\big\} &= -\epsilon^{AB}\delta_{m+n+1}+\frac{3G}{\ell}\epsilon^{AB}\sum_{k\neq 0}\frac{1}{k}\epsilon_{CD}\eta^C_{n+k}\eta^D_{m-k}-\frac{2G}{\ell}\sum_{k\neq 0}\frac{1}{k}\eta^{(A}_{n+k}\eta^{B)}_{m-k},\label{Dbrack1}\\
\big\{{\bar\eta}^{A'}_n,\bar{\eta}^{B'}_m\big\} &= -\bar{\epsilon}^{A'B'}\delta_{m+n+1}+\frac{3G}{\ell}\bar{\epsilon}^{A'B'}\sum_{k\neq 0}\frac{1}{k}\bar{\epsilon}_{C'D'}\bar{\eta}^{C'}_{n+k}\bar{\eta}^{D'}_{m-k}-\frac{2G}{\ell}\sum_{k\neq 0}\frac{1}{k}\bar{\eta}^{(A'}_{n+k}\bar{\eta}^{B')}_{m-k}.\label{Dbrack2}
\end{subalign}
In the $\ell\rightarrow\infty$ limit of vanishing cosmological constant, we are back to the ordinary Heisenberg commutation relations \eref{Hpoiss}, see also \cite{Wieland:2018ymr}.
\subsection{Witt algebra of diffeomorphism charges}
\noindent Finally, let us compute the Poisson commutation relations for the boundary charges \eref{HTdef} that generate conformal boundary diffeomorphisms on the covariant phase space. We will see, in fact, that the algebra defines a representation of the Viraso algebra with vanishing central charge.

The boundary charge \eref{HTdef}, is determined by the shear component $\sigma\propto T_{ab}\partial^a_z\partial^b_z$ of the boundary stress energy tensor. As we have seen in above, the shear $\sigma=2\partial^a_z\partial^b_z \nabla_{(a}n_{b)}$ of the normal vector $n^a\perp T\mathcal{B}$ to the boundary can be expressed in terms of the holomorphic spin coefficient $\xi_A\mathcal{D}\xi^A$, see \eref{HTdef}. Since the $SL(2,\C)$ connection is given by $A=g^{-1}\di g$, and $\eta^A=\ou{g}{A}{B}\xi^B$ is holomoprhic,  the shear of $n^a$ is now simply given by the holomorphic function $\eta_A\partial_z \eta^A=\xi_A\partial^a_z\mathcal{D}_a\xi^A$, which is integrated over a cross-section $\mathcal{C}$ to obtain the quasi-local energy $H[N]$.

In the last section, we introduced an extended phase space $\mathcal{P}^\uparrow_\Sigma$ of field configurations $(\eta^A(z),$ $\ou{h}{A}{B}(z),$ $\ou{U}{A}{B}(z_o^\pm))$, which is equipped with the symplectic structure \eref{Omdef3}. On the extended phase space, the $SL(2,\C)$ element $\ou{h}{A}{B}(z)$, which has a branch cut along the negative real axis, and $\eta^A(z)$, which is holomorphic in $\C-\{0\}$, are functionally independent.\footnote{The group element $h(z)$ has a branch cut, but $\partial_z h^{-1}$ is holomorphic in $\C-\{0\}$.}  The physical phase space  is obtained by imposing two kinds of constraints: the infinite tower of Kac\,--\,Moody constraints \eref{Jcons}, which are local in $z$, and the non-local condition on the monodromy \eref{Cdef}. The resulting infinite-dimensional constraint hypsersurface is equipped with the Dirac bracket \eref{Dbrack}, which turns it into a phase space. 

Since we are working on this extended phase space, we now need to lift the diffeomorphism charges \eref{HTdef} onto $\mathcal{P}^\uparrow_\Sigma$, and we achieve this by introducing the following complex-valued charge 
\begin{equation}
L[N] = \frac{\I}{\sqrt{2}}\oint_{\mathcal{C}}\di z\, N\Big[\eta_A\partial_z\eta^A-\frac{\ell}{8\pi G}\mathrm{Tr}\big(\partial_zh\partial_zh^{-1}\big)\Big],\label{Ldef}
\end{equation}
where the smearing function $N(z)$, which defines the $z$-component of the conformal Killing vector at the boundary, is holomorphic in $\C-\{0\}$. Since $\mathrm{Tr}(\partial_zh \partial h^{-1})$ vanishes on the constraint hypersurface,\footnote{N.B.\ $\eta_A\eta^A =\epsilon_{BA}\eta^B\eta^A =- \epsilon_{AB}\eta^B\eta^A=0$, hence $\mathrm{Tr}(\partial_zh\partial_z h^{-1})=\partial_z\ou{h}{A}{B}\partial_z\ou{[h^{-1}]}{B}{A}\approx -\ell^2/(4\pi G)^2\eta^A\eta_B\eta^B\eta_A=0$.} the quasi-local Hamiltonian  \eref{HTdef} is given by the real part of $L[N]$, 
\begin{equation}
H[N] \approx L[N]+\CC,
\end{equation}
where $\approx$ denotes equality up to terms that vanish on the constraint hypersurface, which is defined by the imposition of both the Kac\,--\,Moody constraints \eref{Jcons} and \eref{Cdef}. To identify the Hamiltonian vector field of $L[N]+\CC$ on the extended phase space, let us first evaluate the functional differential,
\begin{equation}
\bbvar{d} L[N] = \I\sqrt{2}\oint_{\mathcal{C}}\di z\,\bigg[\bbvar{d}\eta_A\Big(N\partial_z\eta^A+\frac{1}{2}\partial_z N\eta^A\Big)-\frac{\ell}{8\pi G}N\,\mathrm{Tr}\Big(\big(\partial_z\bbvar{d} h\big)\partial_z h^{-1}\Big)\bigg].\label{dLN}
\end{equation}
Next, we define the following vector field $X_N\in T\mathcal{P}_\Sigma^\uparrow$ on the extended phase space,
\begin{subalign}
\ou{\big[h^{-1}X_N[h]\big]}{A}{B} &=\sqrt{2}N\ou{\big[h^{-1}\partial_z h\big]}{A}{B},\label{Hactn1}\\
\ou{\big[\bar{h}^{-1}X_N[\bar{h}]\big]}{A'}{B'} &=\sqrt{2}\bar{N}\ou{\big[\bar{h}^{-1}\partial_z \bar{h}\big]}{A'}{B'},\label{Hactn2}\\
X_N[\eta^A]&=\sqrt{2}\Big(N\partial_z\eta^A+\frac{1}{2}(\partial_zN)\eta^A\Big),\label{Hactn3}\\
X_N[\bar{\eta}^{A'}]&=\sqrt{2}\Big(\bar{N}\partial_{\bar{z}}\bar{\eta}^{A'}+\frac{1}{2}(\partial_{\bar{z}}{\bar{N}})\bar{\eta}^{A'}\Big).\label{Hactn4}
\end{subalign}
The action of the vector field $X_N$ (as a functional derivative) on the  $SU(2)$ elements at the marked boundary points $\partial\mathcal{C}=\{z_o^+\}\cup\{z_o^-\}$ of the contour $\mathcal{C}\subset\mathcal{B}$ is arbitrary, because there are the residual and rigid $SU(2)$ gauge transformations \eref{Yvec} that always allow us to set $X_N[U]$ to zero by sending $X_N$ into some $X_N-\lambda^i_NY_i$. We may define, therefore, without any loss of generality that
\begin{equation}
X_N[U(z_o^\pm)]=0.
\end{equation}
If we now contract $\Omega_\Sigma^\uparrow$ with both $X_N$ and a second arbitrary field variation $\delta\in T\mathcal{P}_\Sigma^\uparrow$, we immediately find
\begin{align}
\nonumber \Omega_\Sigma^\uparrow(X_N,\delta)=\I\sqrt{2}&\oint_{\mathcal{C}}\bigg[\di z\Big(N\partial_z\eta_A+\frac{1}{2}\partial_zN\eta_A\Big)\delta\eta^A-\frac{\ell}{8\pi G}N\,\mathrm{Tr}\big(h^{-1}\partial_z h\di(h^{-1}\delta h)\big)-\CC\bigg]+\\
&-\frac{\I\sqrt{2}\,\ell}{16\pi G}\bigg[N \Tr\left(h^{-1}\delta h(h^{-1}\partial_zh)\right)-N\Tr\left(h^{-1}\partial_z h \delta[U]U^{-1}\right)-\CC\bigg]\bigg|_{\partial\mathcal{C}}.\label{XNinOm}
\end{align}
Using $\mathrm{Tr}(\tau_i\tau_j\tau_k)=-\frac{1}{4}\epsilon_{ijk}$, we have
\begin{equation}
\mathrm{Tr}\big(h^{-1}\partial_z h(h^{-1}\partial_zh)(h^{-1}\delta h)\big)=0,
\end{equation}
such that the first line of equation \eref{XNinOm} simplifies to give the differential $-\delta L[N]+\mathrm{cc}$. If the  contribution from the marked boundary points vanishes in \eref{XNinOm}, the vector field $X_N$ will be the Hamiltonian vector field of the quasi-local Hamiltonian  $L[N]+\CC\approx H[N]$. This happens as soon as we restrict ourselves to field variations $\delta\in T\mathcal{P}^\uparrow$ that are tangential to the $C=0$ constraint hypersurface, i.e.\ $\delta[C]=0$. In fact, if we reintroduce $g= hU$, we have
\begin{equation}
\mathrm{Tr}\big(h^{-1}\delta h(h^{-1}\partial_z h)\big) = \mathrm{Tr}\big(g^{-1}\delta(g) U^{-1}(h^{-1}\partial_z h)U\big) +\mathrm{Tr}\big(h^{-1}\partial_z h\,\delta[U]h^{-1}\big).\label{XNinOm2}
\end{equation}
On the $C=0$ constraint hypersurface, which is defined by \eref{Cdef}, the first term is continuous across the branch cut, and the second term on the right hand side of \eref{XNinOm2} will cancel against the last term in the second line of \eref{XNinOm}. For any field variation $\delta$ that preserves the constraint on the monodromy, the boundary terms  cancel, and we find that $X_N$ is indeed the Hamiltonian vector field of $L[N]+\CC$ on the physical phase space,
\begin{equation}
\Omega_\Sigma^\uparrow(X_N,\delta) \approx -\delta L[N]+\CC,\quad\forall\delta\in T\mathcal{P}_\Sigma^\uparrow: \delta[C]\approx0,\label{Hamvec}
\end{equation}
where $\approx$ denotes equality up to terms that vanish if the $C=0$ constraint on the monodromy \eref{Cdef} is satisfied.  A particular example of such a field variation $\delta$ is given by the vector field $X_N$ itself: although $h(z)\in SL(2,\C)$ may have a branch cut, the extended phase space $\mathcal{P}_\Sigma^\uparrow$ contains only such configurations where $\partial_z hh^{-1}$ is continuous across the branch cut. This implies
\begin{equation}
X_N[C]=\sqrt{2} N(z_o^+)(\partial_zhh^{-1})\big|_{z_o^+} C-\sqrt{2} N(z_o^-)(\partial_zhh^{-1})\big|_{z_o^-} C\approx0.\label{XCons}
\end{equation}
Therefore, $X_N$ is a vector field in $\mathcal{P}_\Sigma^\uparrow$ that lies tangential to the $C=0$ hypersurface. In addition, the vector field $X_N$ preserves the Kac\,--\,Moody constraints \eref{Jdef},
\begin{equation}
X_N[J^i_n]\approx0.
\end{equation}
We have thus shown that the vector field $X_N\in T\mathcal{P}_\Sigma^\uparrow$ lies  tangential to the entire constraint hypersurface $\mathcal{P}_\Sigma=\big\{p=[\eta^A(z),\ou{h}{A}{B}(z),\ou{U}{A}{B}(z^\pm_o)]:C(p)=0=J^i_n(p)\big\}$. On the physical phase space, equation \eref{Hamvec} implies that the Hamiltonian vector field of $H[N]$ is given by $X_N$.

Having identified the Hamiltonian vector fields, we can now immediately evaluate the corresponding Poisson algebra. We contract the symplectic two-form with any two such vector fields and obtain
\begin{align}\nonumber
\Omega_{\Sigma}^\uparrow (X_N,X_{M}) & = \I\oint_{\mathcal{C}}\Big[\di z\Big(N\partial_z\eta_A+\frac{1}{2}\partial_z N\eta_A\Big)\Big(M\partial_z\eta^A+\frac{1}{2}\partial_z M\eta^A\Big)+\\\nonumber
&\hspace{5em}-\frac{\ell}{8\pi G}N\Tr\left(h^{-1}\partial_zh\,\di \big(M h^{-1}\partial_z h\big)\right)-(N\leftrightarrow M)\Big]+\CC=\\\nonumber
&=-\I\oint_{\mathcal{C}}\di z\,\left(N\partial_z M-M\partial_z N\right)\Big[\eta_A\partial_z\eta^A-\frac{\ell}{8\pi G}\Tr\left(\partial_zh\partial_zh^{-1}\right)\Big]+\CC=\\
&=-H\big[[N,M]\big],
\end{align}
where we defined the Lie bracket\footnote{The prefactor of $\sqrt{2}$ is a consequence of our conventions for the conformal Killing vector $t^a_N:t_N^a = N\bar{m}^a+\CC$, with $m^a =\sqrt{2}\partial^a_{\bar{z}}$ and $[t_N,t_M]^a=t_{[N,M]}^a$.}
\begin{equation}
[N,M]=\sqrt{2}\big(N\partial_zM-M\partial_zN\big).
\end{equation}
The constraint hypersurface is equipped with a natural symplectic form, which is given by the pull-back of \eref{Omdef3} from the auxiliary phase space $\mathcal{P}_\Sigma^\uparrow$ back to $\mathcal{P}_\Sigma$. The corresponding Poisson brackets on the constraint hypersurface are given by the Dirac bracket, \eref{Dbrack}. On the $C=0$ constraint hypersurface, the Hamiltonian vector field of $H[N]$ is given by $X_N$, which preserves all the Kac\,--\,Moody constraints,\begin{equation}
\big\{H[N],J^i_n\big\}^\uparrow\Big|_{C=0}=0.
\end{equation}
This implies that the commutation relation for the generators $H[N]$ are unaffected by the presence of the Dirac bracket,
\begin{align}\nonumber
\big\{H[N],H[M]\big\}\Big|_{C=0}&=\big\{H[N],H[M]\big\}^\uparrow\Big|_{C=0}=\Omega_{\Sigma}^\uparrow (X_N,X_{M})\Big|_{C=0}=-H\big[[N,M]\big]\Big|_{C=0},
\end{align}
where $\{\cdot,\cdot\}^\uparrow$ denotes the Poisson brackets on the extended phase space. If we  define  the usual Virasoro charges, 
\begin{equation}
L_n =\frac{1}{\sqrt{2}}L[z^{n+1}].
\end{equation}
we immediately find two copies of the Virasoro algebra with vanishing central charge
\begin{subalign}
\big\{L_n,L_m\big\} \approx (n-m)L_{n+m},\label{Vir1}\\
\big\{\bar{L}_n,\bar{L}_m\big\}\approx (n-m)\bar{L}_{n+m},\label{Vir2}
\end{subalign}
where $\approx$ denotes again terms that vanish provided the closure constraint \eref{Cdef} for the monodromy is satisfied.

It is now instructive to evaluate the Virasoro charges for the simplest non-trivial classical solution, namely Bryant's curved catenoid cousins \cite{AST_1987__154-155__321_0,Bobenko2009}. In fact, we have identified an entire one-parameter family $\{\mathcal{B}_a\}_{0\leq a < 1/\sqrt{2}}$ of such CMC-1 surfaces in Euclidean $\AdS$, see \eref{etax}, and \eref{rhodef1}. Any such catenoid cousin is now characterised by the holomorphic boundary spinor, 
\begin{equation}
\eta^A[{\mathcal{B}_a}] = \I a\sqrt{\frac{\ell}{8\pi G}}\begin{pmatrix}1\\-z^{-1}\end{pmatrix}.
\end{equation}
For any such configuration all but one of the quasi-local boundary charges vanish,
\begin{equation}
L_n[\mathcal{B}_a]=\frac{\I}{2}\oint \di z\, z^{n+1}\eta_A\partial_z\eta^A\Big|_{\mathcal{B}_a}=\frac{L a^2}{8 G}\delta_n.
\end{equation}
The limit to the asymptotic boundary is the limit $a\rightarrow 1/\sqrt{2}$. In this limit, we recover the asymptotic value of the $\AdS$ vacuum energy,
\begin{equation}
L_n[\partial \AdS]:= \lim_{a\rightarrow\frac{1}{\sqrt{2}}}L_n[\mathcal{B}_a] = \frac{\ell}{16 G}\delta_{n}.\label{Lbndry}
\end{equation}

Before we proceed, let us briefly summarise the results of this section. To compute the Poisson commutation relations between the quasi-local boundary observables $H[N]=\oint dv^aT_{ab}t^b_N$, we found it useful to work on an extended phase space $\mathcal{P}_\Sigma^\uparrow$, where the  functional dependence between the holomorphic boundary spinors $\eta^A(z)$ and the holomorphic $SL(2,\C)$ elements $\ou{h}{A}{B}(z)$ is removed such that $\eta^A(z)$ and $\ou{h}{A}{B}(z)$ can be treated as independent coordinates on the extended phase space. The physical phase space is obtained by imposing the conditions that reestablish the functional dependence between $\eta^A(z)$ and $\ou{h}{A}{B}(z)$, namely by imposing the Kac\,--\,Moody constraints \eref{Jcons}, and the non-local closure constraint \eref{Cdef} on the monodromy. By introducing the Virasoro generators $L[N]$, we then lifted the Hamiltonian charges $H[N]$ onto this extended phase space. Next, we found specific vector fields $X_N\in T\mathcal{P}_\Sigma^\uparrow$ that lie tangential to the solution space of the constraints \eref{Jdef} and \eref{Cdef}, and coincide on the constraint hypsersurface with the Hamiltonian vector fields of $H[N]$, see \eref{Hamvec}. Having identified the Hamiltonian vector fields of $H[N]$, we  then found the corresponding Poisson (Dirac) brackets on the solution space of the constraint equations. We recovered two copies of the Virasoro algebra with vanishing central charge.

\section{Entropy and partition function}\label{sec5}
\noindent Finally, a few remarks on quantum gravity and black holes. Our main goal in this section is to gather some evidence that the conformal boundary spinors $\xi^A$ provide a microscopic explanation for black hole entropy. Our discussion relies on the observation due to Strominger \cite{Strominger:1997eq} that the Bekenstein\,--\,Hawking entropy for a three-dimensional black hole has the same algebraic structure as the Cardy formula \cite{Cardy:1986ie} for a two-dimensional conformal field theory,
\begin{equation}
S(M,J) = \frac{2\pi \rho(r_+)}{4 G}=\pi\sqrt{\frac{\ell}{2G}}\,\left(\sqrt{M\ell+J}+\sqrt{M\ell-J}\right),\label{BHentropy}
\end{equation}
where $M$ and $J$ are the mass and angular momentum of the BTZ black hole. The Cardy formula holds for a large class of boundary CFTs, which makes the argument robust, but it does not tell us much about the field content of the boundary CFT. In the following, we would like to discuss this issue from the perspective of the boundary modes $\xi^A$. To this goal, let us consider first the bulk plus boundary path integral,
\begin{equation}
X(\tau,\bar{\tau}) = \int_{\mathcal{M_\tau}} \boldsymbol{\mathcal{D}}[e,\omega]\,\boldsymbol{\mathcal{D}}[\xi]\,\E^{\I S[e,\omega|\xi]},
\end{equation}
where $\mathcal{M}_\tau$ is a solid torus, which is characterised by the modular parameter\footnote{The extra imaginary unit infront of $(\beta+\I\varphi)$ has to do with the fact that we are considering an oscillatory integral.} $\tau=\frac{\I}{2\pi}(\beta+\I\varphi)$ that encodes the periodicity $z\sim z\, \E^{\beta+\I\varphi}$ on the complex plane $\C-\{0\}$. Notice that the exponent is imaginary, because the Euclidean  bulk plus boundary action \eref{actndef} is real. We are considering, therefore, an oscillatory path integral, which also underlies the Ponzano\,--\,Regge and Turaev\,--\,Viro spinfoam amplitudes and their generalisations to four dimensions \cite{TURAEV1992865,ponzanoregge,alexreview, Barrett:2008wh,Freidel:2007py,LQGvertexfinite,Engle:2007uq}. 

The integral over the the triad $e^i$ and the $SU(2)$ connection $\omega^i$ in the interior is redundant, because there are no radiative degrees of freedom in the bulk (the $\omega^i$ and $e^i$ directions lie tangential to the gauge orbits). We are thus left with the path integral over the boundary spinors alone, which defines a Virasoro character\footnote{We have absorbed a potential vacuum energy back into the definition of $L_0$.}
\begin{equation}
X(\tau,\bar{\tau})= \Tr\left(\E^{-2\pi\tau L_0}\E^{2\pi\bar{\tau}\bar{L}_0}\right)=\mathrm{Tr}\Big(\E^{-\I\beta H+\varphi J}\Big).\label{Lcharac}
\end{equation}
Only  those  states will contribute to this trace that satisfy the infinitely many Kac\,--\,Moody constraints $J^i_n\approx 0$, which are imposed  via the Dirac bracket at the classical level. In addition, we have to impose also the closure constraint on the monodromy \eref{Cdef}. The closure constraint mixes the holomorphic and anti-holomorphic sectors of the theory and we cannot assume, therefore, that the character factorises, $X(\tau,\bar{\tau})\neq X(\tau)\bar{X}(\bar\tau)$.

The Hamiltonian $H$, which generates translations along the radial $|z|$-direction, and the angular momentum $J$ are the real and imaginary\footnote{The Euclidean BTZ black hole solution is characterised by an imaginary spin $J$, and a positive mass $M\geq \ell\,|J|$.}  part of the Virasoro generators $L_0$ and $\bar{L}_0$,
\begin{subalign}
H&=L_0+\bar{L}_0,\\
J&=L_0-\bar{L}_0.
\end{subalign}
In our case, the Virasoro generators will satisfy the reality conditions
\begin{equation}
L_n^\dagger = \bar{L}_n,
\end{equation}
where $L_0^\dagger$ denotes the Hermitian conjugate with respect to the Hilbert space inner product. Notice, that there is no reason a priori for the Virasoro generators \eref{Ldef} to satisfy the more familiar adjointness relations $L_{-n}= L^\dagger_n$ (and $\bar{L}^\dagger_n=\bar{L}_{-n}$) that underpin conventional unitary CFTs.\footnote{We could insist to use a bilinear form $(\cdot,\cdot)$ such that $(L_{-n}[\cdot],\cdot)=(\cdot, L_n[\cdot])$, but then the requirement of positivity for $(\cdot,\cdot)$ must be dropped.} %We must expect, in fact, that the boundary field theory for the boundary spinors $\xi^A$ defines a non-unitary CFT, because the boundary action contains the kinetic term for the $\beta$-$\gamma$-ghosts of string theory, which are now coupled to  a quartic potential $V(\beta,\gamma)\simeq|\beta|^4+|\gamma|^4$ and an interaction term that couples the boundary modes to the $SU(2)$ spin connection from the bulk.\footnote{The interaction term comes from the covariant derivative, the relation to the $\beta$-$\gamma$ ghosts is provided by the identification $(\xi_0,\xi_1)=(\beta,\gamma)$.} 

If the boundary spinors $\xi^A$ are the origin of  black hole entropy, we should be then able to compute the micro-canonical entropy 
\begin{equation}
S(\Delta,\bar{\Delta}) = \log \Omega(\Delta,\bar{\Delta}),
\end{equation}
where $\Omega(\Delta,\bar{\Delta})$ is the number of boundary states that have energy $E$ and spin $L$, with $\Delta = E+\I L$ denoting the (complex) eigenvalue of $L_0= H+J$. The degeneracy of $L_0$ can be then calculated by an averaging procedure  \cite{Carlip:2005zn,Maloney:2007ud}, namely by taking the Laplace transform of the character,
\begin{equation}
\Omega(\Delta,\bar{\Delta}) = \operatorname{Tr}\left(\delta(L_0-\Delta)\delta(\bar{L}_0-\bar{\Delta})\right)=\frac{1}{(2\pi)^2}{\int_{\gamma^\ast}}\!\!\di \tau^\ast\!{\int_{\gamma}}\di\tau \,X(\tau,\tau^\ast) \E^{2\pi\tau\Delta}\E^{-2\pi\tau^\ast\bar{\Delta}},
\end{equation}
where we have analytically continued $X(\tau,\bar{\tau})$ into an analytic function $X(\tau,\tau^\ast)$ of two complex numbers $(\tau,\tau^\ast)$. The paths $\gamma(t)=\tau(t)$ and $\gamma^\ast(t)=-\bar{\tau}(t)$ are chosen such that the integral converges. Since we do not know the spectrum of $L_0$, we can now only proceed at a formal level.  Suppose, therefore, that for some given configuration $(\Delta,\bar{\Delta})$ the integral converges in both $\tau$ and $\tau^\ast$ and that the main contribution to the integral comes from a single saddle point\footnote{If $\tau_o$ is such a saddle point for the integral over $\tau$, then $-\bar{\tau}_o$ will be the saddle point for the $\tau_o^\ast$ integral.} $\tau_o$ (res.\ $\tau^\ast_o$).  The relation between the entropy $S(\Delta,\bar{\Delta})$, and the inverse temperature $\tau_o$ is then given by the usual saddle point equations 
\begin{subalign}
 2\pi\Delta&=-\frac{\partial\log X}{\partial\tau}\Big|_{(\tau_o,-\bar{\tau}_o)} ,\\
S(\Delta,\bar{\Delta}) &\approx \log X(\tau_o,-\bar{\tau}_o)+2\pi\tau_o\Delta + 2\pi \bar{\tau}_o\bar{\Delta}.
\end{subalign}
We can now formally continue to derive a version of the Cardy formula: since the modular $S$-transformation $\tau\rightarrow -\tau^{-1}$ defines the same torus, we expect that the Virasoro character of the boundary field theory is invariant under these large diffeomorphisms. Let us then also assume that for large temperature $|\tau_o|\rightarrow 0$ the integral over the oscillating trace is dominated by a single semi-classical (coherent) state $|\Omega\rangle$, 
\begin{align}
X(\tau,\tau^\ast)\approx \E^{{\frac{2\pi}{\tau_o}\langle \Omega|L_0|\Omega\rangle}}\E^{-{\frac{2\pi}{\tau^\ast_o}\langle \Omega|\bar{L}_o|\Omega\rangle}}.
\end{align}
 If  such a state $|\Omega\rangle$ exists, we can immediately perform the Legendre transformation from $X(\tau,\tau^\ast)$ to $S(\Delta,\bar{\Delta})$ and obtain,
\begin{subalign}
\tau_o&=\sqrt{\frac{\Delta}{\langle \Omega|L_0|\Omega\rangle}},\\
S(\Delta,\bar{\Delta}) & \approx 4\pi\sqrt{\Delta \langle \Omega|L_0|\Omega\rangle}+\CC\label{cardy}
\end{subalign}
Notice that we do not require that $|\Omega\rangle$ is an eigenstate of $L_0$. In fact, for a non-unitary CFT $L_0+\bar{L}_0$ may be unbounded from below and may have no normalisable eigenstates in the Fock space of the boundary CFT.\footnote{In the $\ell\rightarrow 0$ limit of vanishing cosmological constant, the quasi-local energy $H=L_0+\bar{L}_0$ turns into a two-mode squeeze operator, which has no normalisable eigenstates on the Hilbert space of the boundary CFT, see \cite{Wieland:2018ymr}.}  

Given these assumptions (namely, (i) modular invariance and the (ii)  existence of a semi-classical state $|\Omega\rangle$ that dominates the partition function at large temperature), we  have a version of Cardy's formula for the boundary CFT, which we expect to be non-unitary. For a generic such coherent state $|\Omega\rangle$ the entropy \eref{cardy} is in violation of the Bekenstein\,--\,Hawking formula \eref{BHentropy}. If, however, the semi-classical state $|\Omega\rangle$, represents the asymptotic boundary, which corresponds to the $a\rightarrow\frac{1}{\sqrt{2}}$ limit of the bulk catenoids (\ref{etax}, \ref{rhodef1}) the situation is different: since the state is assumed to be semi-classical, we would then recover the classical values for the Virasoro generators (at least to leading order in $\hbar$). In other words,
\begin{equation}
\langle \Omega|L_0|\Omega\rangle = \frac{\ell}{16 G},\label{vac}
\end{equation}
see \eref{Lbndry}. Equation \eref{cardy} together with \eref{vac} would then reproduce the Bekenstein\,--\,Hawking entropy, for mass $M$ and imaginary spin $J$ that are now determined by the real and imaginary parts of $\Delta=M\ell+J$.

The key open task to make this argument robust is to show that there exists a  coherent state $|\Omega\rangle$ that represents the asymptotic $\AdS$ boundary and dominates the Virasoro character at high temperature. This task is not unfeasible, because there has been a lot of progress in non-perturbative quantum general relativity to construct such coherent boundary states using the spin network representation, which would provide a lattice regularisation  of the boundary CFT, see for instance \cite{Bianchi:2012ev,Bianchi:2016tmw,Freidel:2010tt,twist3,Girelli:2005ii} and references therein. In fact, using a coherent spin network for the quantum states in the bulk, we will have a coherent boundary state $\Omega$ that will be now only supported in a finite  number of punctures, i.e.\ $\langle\xi|\Omega\rangle=\Omega[\xi^A(z_1),\xi^A(z_2),\dots]$, with every such puncture representing a gravitational Wilson line that ends at the boundary \cite{Wieland:2017cmf,Wieland:2018ymr}. Introducing a UV cutoff for the mode expansion %\footnote{Short distances with respect to the auxiliary metric $q_{ab}$ of the boundary CFT do not necessarily represent short distances with respect to the physical metric $h_{ab}$, because the conformal factor is a composite field of the CFT. } 
of the boundary CFT, one can then map the  Hilbert space of $N$ such punctures back into the Hilbert space of the boundary CFT in the continuum, which would then allow us to test the validity of the approximation \eref{cardy}, see  \cite{Wieland:2017cmf,Wieland:2018ymr} and \cite{Kempf:2010rx} for related developments based on the sampling theorem. 
\section{Summary and discussion}
\noindent In this paper, we established the quasi-local Hamiltonian formulation of three-dimensional Euclidean gravity ($\Lambda=-1/\ell^2$) with conformal boundary conditions. The conformal class of the induced metric at the boundary is fixed, but there are no restrictions on the variations of the conformal factor. Instead, there are constraints on the canonically conjugate variable to $\log\Omega$, which is the trace $K=\nabla_an^a$ of the extrinsic curvature. The specific value $K=2/\ell$ is preferred geometrically, because the solution space of this specific class of conformal boundary conditions can be coordinatised in terms of holomorphic maps from Riemann surfaces into the spin bundle \cite{AST_1987__154-155__321_0, Bobenko2009} over hyperbolic space. To impose the conformal boundary conditions at a Hamiltonian level, we found it then useful to work on an extended phase space \cite{Wieland:2017cmf,Donnelly:2016auv}, where there are additional boundary degrees of freedom that turn these holomorphic maps into dynamical boundary fields, whose Euclidean time evolution is governed by the Noether charges \eref{HTdef}. %The conformal boundary conditions \eref{bndrycond} are then imposed dynamically by adding the boundary terms \eref{actndef} to the triadic Palatini action in the bulk \eref{bulkactn}. At the saddle points of the coupled bulk plus boundary theory, we recovered the corresponding bulk plus boundary equations of motion, namely the usual field equations in the interior (\ref{TEOM}, \ref{FEOM}) in addition to the boundary field equations (\ref{EOM2}, \ref{glucond}) that impose the conformal boundary conditions \eref{bndrycond} along the cylindrical boundary.

After having introduced the appropriate counter terms to the triadic Palatini action \eref{actndef}, we studied the phase space and the gauge symmetries of the bulk plus boundary system. Simultaneous $SU(2)$ frame rotations of the bulk plus boundary fields are unphysical gauge directions. For diffeomorphisms, the situation is different: large diffeomorphism are physical \cite{Balachandran:1994up,Carlip:2005zn}. A preferred class of such large diffeomorphisms is given by those specific bulk diffeomorphisms that preserve the conformal boundary conditions. The corresponding conserved Noether charges are the Virasoro generators \eref{HTdef}. 

Finally, we computed the Poisson commutation relations for the holomorphic boundary spinors, and found a one-parameter family of deformations of the classical Heisenberg algebra: $\{\eta^A_n,\eta^B_m\}=-\epsilon^{AB}\delta_{m+n+1}+\sum_{rs} r^{AnBm}_{CrDs}\eta^C_{r}\eta^D_{s}$. The infinite-dimesnional matrix $r^{AnBm}_{CrDs}\sim G/\ell$ that controls the strength of this deformation disappears in the $\ell\rightarrow\infty$ limit of vanishing cosmological constant. The geometrical origin of this deformation can be traced back to the  Kac\,--\,Moody constraints \eref{Jcons}, which are second-class. In fact, $r^{AnBm}_{CrDs}$ is simply the inverse of the Dirac matrix $\{J^i_n,J^k_m\}$ of the second-class constraints. Besides the Kac\,--\,Moody constraints there is  a small number of residual first-class constraints, namely the zero mode $J^i_0$ of the Kac\,--\,Moody charges in addition to the deformed closure constraint \eref{gauss} that entangles the holomorphic and anti-holomorphic sectors of the boundary field theory.
 
 From the perspective of the spin network representation of quantum general relativity, the field content of the boundary CFT should be no surprise. In loop quantum gravity (LQG), the quantum states of the geometry are constructed by successively exciting gravitational Wilson lines for the spin connection.\footnote{The underlying diffeomorphism invariant Ashtekar\,--\,Lewandowski vacuum is a state that represents no geometry at all \cite{ALvacuum}. More recently, dual vacua have been proposed that are peaked in the conjugate variables: the metric is widely spread, but the conjugate momentum, which encodes the extrinsic curvature is sharply peaked, see \cite{Bahr:2015bra,Dittrich:2014ala}.} The introduction of a boundary breaks these Wilson lines apart, and excites a distributional surface charge, namely a boundary spinor, at the puncture. In $2+1$ spacetime dimensions, the partition function for these gravitational boundary modes is given by the evaluation of boundary spin networks against the Ponzano\,--\,Regge spinfoam amplitudes. The resulting spin network evaluation defines a large class of $1+1$-dimensional statistical model \cite{Bonzom:2015ans,Dittrich:2018xuk,Dittrich:2017hnl,Dittrich:2017rvb}. The Heisenberg XYZ spin chain is an example for such a statistical model in $1+1$ dimensions, which corresponds to the massive Thirring model in the continuum.  The results of this paper strengthen these dualities   from the opposite direction, namely by starting from a $1+1$ dimensional boundary field theory for conformal boundary conditions in the continuum.
 
The main part of the paper dealt with the classical theory. In the last section, we discussed  the physical relevance of our results in the context of those proposals that derive the entropy of black holes from the Cardy formula. Our discussion closely followed Strominger's original proposal, but there are a few unusual features. First of all, we found that there is no central charge among the Poisson brackets of the Virasoro generators, see \eref{Vir1} and \eref{Vir2}. In our opinion, this is a strong indication that the underlying boundary CFT is non-unitary. This observation is further supported by the structure of the boundary action, \eref{actndef}. If we isolate the spin up and down components of the boundary spinor and introduce component functions $\beta$ and $\gamma$, such that $\xi_A=(\beta,\gamma)$, we will find that the boundary action \eref{actndef} turns into the action for the $\beta$-$\gamma$ ghosts of string theory, with a quartic potential and a minimal coupling to the spin connection from the bulk (the boundary CFT resembles, therefore, the Thirring model \cite{THIRRING195891}, but with a kinetic term, which is now borrowed from the $\beta$-$\gamma$ theory). Yet the statistics is different, since $\xi^A$ must be bosonic.\footnote{At least at the classical level: if $\xi^A$ were Grassmann-valued, we would have $\xi^A(z)\xi^B(z)\propto\epsilon^{AB}$ and the gluing condition \eref{glucond} would imply that the triad at the boundary vanishes $\varphi^\ast_{\mathcal{B}}e_i\propto\sigma_{ABi}\epsilon^{AB}=-\mathrm{Tr}(\sigma_i)=0$.} The main difficulty in quantising such a theory is that  $H=L_0+\bar{L}_0$ is not manifestly positive. This becomes explicit in the $\ell\rightarrow\infty$ limit of vanishing cosmological constant, where $H=L_0+\bar{L}_0$ is a two-mode squeeze operator $H\sim \sum_n (2n+1)(a_nb_n+a^\dagger_n b^\dagger_n)$, see \cite{Wieland:2018ymr}. We expect that these features survive for $\Lambda\neq 0$ and that the Hamiltonian will have a similar spectrum, such that the exponentials of the Hamiltonian vector fields of the supermomentum generators $P_n=\ell(L_n+\bar{L}_n)$ would not preserve the original Fock space, but map it into a unitarily inequivalent superselection sector. 
\vspace{-0.6em}

\paragraph{- Acknowledgments} We would like to thank Bianca Dittrich, Laurent Freidel, Florian Girelli and Simone Speziale for many enlightening discussions. In addition, we would like to thank Perimeter Institute for the opportunities provided by the undergraduate theoretical physics summer program. Research at Perimeter Institute is supported by the Government of Canada through the Department of Innovation, Science and Economic Development and by the Province of Ontario through the Ministry of Research and Innovation.
%\section*{References}
\providecommand{\href}[2]{#2}\begingroup\raggedright\endgroup

\end{document}